\def\K{{\;\text{K}}}
\def\ns{{\;\text{ns}}}

\def\Pc{{P_{\text{qp-pair}}}}
\def\Ps{{P_{\text{pair}}}}
\def\Pqp{{P_{\text{qp}}}}

\documentclass[aps,prappl,twocolumn,groupedaddress,showkeys,showpacs,superscriptaddress,floatfix]{revtex4-1}
\usepackage{epsfig}
\usepackage{multirow}
\usepackage{amsmath, amssymb,mathtools}
\usepackage{color}
\usepackage{graphicx}
\usepackage{dcolumn}
\usepackage{bm}
\usepackage{subfigure}
\usepackage[utf8]{inputenc}
\usepackage[T1]{fontenc}
\usepackage{tikz}
\usepackage{soul}

\begin{document}

\title{Solitonic thermal transport in a current biased long Josephson junction}

\author{Claudio Guarcello}
\email{claudio.guarcello@nano.cnr.it}
\affiliation{NEST, Istituto Nanoscienze-CNR and Scuola Normale Superiore, Piazza San Silvestro 12, I-56127 Pisa, Italy}
\author{Paolo Solinas}
\affiliation{SPIN-CNR, Via Dodecaneso 33, 16146 Genova, Italy}
\author{Alessandro Braggio}
\affiliation{NEST, Istituto Nanoscienze-CNR and Scuola Normale Superiore, Piazza San Silvestro 12, I-56127 Pisa, Italy}
\author{Francesco Giazotto}
\affiliation{NEST, Istituto Nanoscienze-CNR and Scuola Normale Superiore, Piazza San Silvestro 12, I-56127 Pisa, Italy}

\date{\today}

\begin{abstract}
We investigate the coherent energy and thermal transport in a temperature-biased long Josephson tunnel junction, when a Josephson vortex, i.e., a soliton, steadily drifts driven by an electric bias current. We demonstrate that thermal transport through the junction can be controlled by the bias current, since it determines the steady-state velocity of the drifting soliton. We study the effects on thermal transport of the damping affecting the soliton dynamics. In fact, a soliton locally influences the power flowing through the junction and can cause the variation of the temperature of the device. When the soliton speed increases approaching its limiting value, i.e., the Swihart velocity, we demonstrate that the soliton-induces thermal effects significantly modify. 
Finally, we discuss how the appropriate material selection of the superconductors forming the junction is essential, since short quasiparticle relaxation times are required to observe fast thermal effects.
\end{abstract}

\maketitle

\section{Introduction}
\label{Sec00}\vskip-0.2cm

Long Josephson junctions (LJJs) are physical systems often used as a framework to explore nonlinear dynamics~\cite{Par93,Ust98}. Nonetheless, coherent thermal transport in this context was explored only recently~\cite{Gua16,Gua18,GuaSolBra18}. In fact, as a temperature gradient is imposed across the device, namely, as the electrodes forming the junction reside at different temperatures, a heat current depending on the configurations of Josephson vortices, i.e., solitons, flows through the device~\cite{Gua18,GuaSolBra18}. The phase-dependent heat current was recently explored theoretically and experimentally in both Josephson junctions (JJs)~\cite{Gia13,Mar14} and superconducting quantum-interference devices (SQUIDs)~\cite{GiaMar12,Gia12}. This phenomenon is the core of the emerging field of phase-coherent caloritronics~\cite{Gia06,MarSol14,ForGia17}, from which fascinating devices, such as heat diodes~\cite{Mar15}, thermal transistors~\cite{For16}, solid-state memories~\cite{Gua17,GuaSol18}, microwave refrigerators~\cite{Sol16}, thermal engines~\cite{Pao18}, thermal routers~\cite{Tim18,Gua18}, and heat amplifier~\cite{Pao17}, were conceived. Recently, it was demonstrated theoretically that a static soliton in a temperature biased long tunnel junction induces a localized warming in one of the electrodes of the device, according to which the application as a fast solitonic thermal router was suggested~\cite{Gua18}. The scenario changes if we consider a time dependent external magnetic field, since more solitons can be excited along the device and the soliton configuration reflects on the temperature profile of the junction, so that every magnetically-excited soliton induces a well-localized temperature peak~\cite{GuaSolBra18}.

In this paper we discuss how an electric bias current affects the thermal transport when a soliton is steadily drifting along the system as driven by the bias current. We demonstrate that the phase-dependent components of the heat current depend on the soliton speed, the latter being also a function of the bias current. Therefore, the dissipationless bias current can be used as a knob to locally modify the thermal transport across the device. Additionally, we study how the damping affecting the soliton dynamics influences the thermal transport. In fact, in the case of a low value of the damping parameter~\cite{Bar82}, we demonstrate that the energy and thermal transport profiles along the junction induced by the soliton significantly change, as the bias current increases.

In the following, we will make two realistic approximations, namely, we write the phase solution representing an electrically driven soliton in a closed simplified form~\cite{PedSam84}, and we consider a soliton traveling at the steady drift velocity~\cite{McL78}.

The paper is organized as follows. In Sec.~\ref{Sec01}, we examine how an electric biasing current generally affects the energy transport in a temperature-biased JJ. In Sec.~\ref{Sec02}, we focus on a soliton moving under the influence of a bias current in a LJJ. In Sec.~\ref{Sec03}, we discuss the behavior of heat currents through the junction as a function of the bias current in the adiabatic limit. In Sec.~\ref{Sec04}, we discuss the adiabatic limit and its implication on material selection in order to find appropriate thermal response timescales. In Sec.~\ref{Sec05}, conclusions are drawn.

\section{Energy transport}
\label{Sec01}\vskip-0.2cm

\begin{figure*}[htbp!!]
\includegraphics[width=0.75\textwidth]{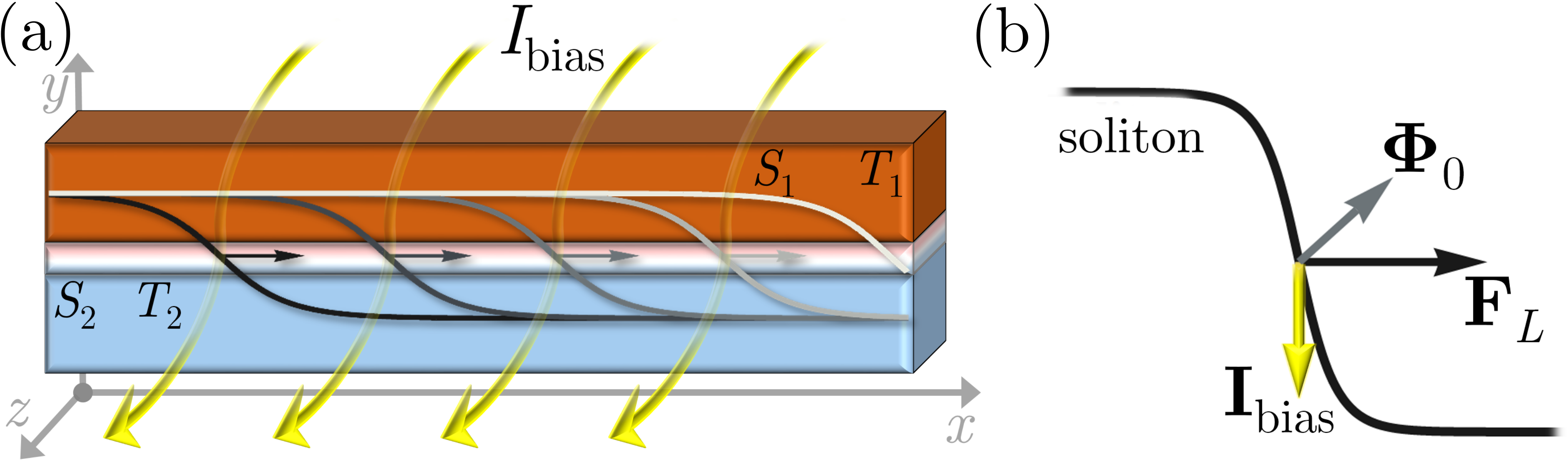}
\caption{(a), A superconductor-insulator-superconductor (SIS) temperature-biased rectangular long Josephson junction (LJJ) driven by an electrical bias current $I_{\text{bias}}$. The temperature $T_i$ of the electrode $S_i$ is also indicated. A soliton drifting due to the bias current is shown. (b) Lorentz force due to a bias current on a soliton. Indeed, a bias current flowing through the junction acts on the soliton with a Lorentz force, $\mathbf{F}_L\propto\mathbf{I}_\text{bias}\times\mathbf{\Phi}_0$, with the direction of $\mathbf{\Phi}_0$ depending on the polarity of the soliton, see Eq.~\eqref{soliton}.}
\label{Fig01}
\end{figure*}

In a current-biased LJJ, the phase difference $\varphi(x,t)$ along the junction in the presence of a soliton can be approximatively expressed as the sum of a dynamics contribution, $\phi(x,t)$, and a static contribution, $\sin^{-1}i_b$, that is~\cite{PedSam84} 
\begin{equation}\label{sumphase}
\varphi(x,t)\simeq\phi(x,t)+\sin^{-1}i_b,
\end{equation}
where the phase shift induced by the bias current,$i_b=I_{\text{bias}}/I_c$ (with $I_c$ being the critical current of the device), is simply added to the soliton solution $\phi$. This dissipationless current can be directly delivered by a current source, or it can coincide with the current circulating in a magnetically driven superconducting ring containing the junction.

In a washboard-like picture~\cite{Bar82}, the term $\sin^{-1}i_b$ in Eq.~\eqref{sumphase} represents the shift of the potential minimum, within which the phase profile $\phi(x,t)$ lies. This shift is induced by the tilting of the potential imposed by the external bias current.

The energy transport in a temperature biased JJ can be written as~\cite{Gut97,GutNat97,Gol13,Vir17,Gua18}
\begin{eqnarray}\label{ptot}\nonumber
P_{\text{tot}}(T_1,T_2,\varphi)&=&\Pqp(T_1,T_2)-\cos\varphi \Pc(T_1,T_2,V)\\
&&+\sin\varphi \Ps(T_1,T_2,V),
\end{eqnarray}
where $V(x,t)=\frac{\Phi_0}{2\pi}\frac{\partial \varphi (x,t) }{\partial t}$ is the local voltage drop ($\Phi_0= h/2e\simeq2\times10^{-15}\; \textup{Wb}$ is the magnetic flux quantum, with $e$ and $h$ being the electron charge and the Planck constant, respectively) and $T_i$ is the temperature of the electrode $S_i$. In Eq.~\eqref{ptot}, $\Pqp$, $\Pc$, and $\Ps$ represent the quasiparticle and the ``anomalous'' contributions to thermal current density flowing through the junction~\cite{Mak65,Fra97,Gut97,Gut98,Gia06,Gol13}, see Appendix~\ref{AppA}. In fact, $\Pqp$ is the heat flux density carried by quasiparticles and represents an incoherent flow of energy through the junction from the hot to the cold electrode. Instead, the ``anomalous'' terms $\Ps$ and $\Pc$ determine the phase-dependent part of the heat current originating from the energy-carrying tunneling processes involving Cooper pairs and recombination/destruction of Cooper pairs on both sides of the junction. We remark that $\Ps$ linearly depends on the voltage drop, so that $\Ps\to0$ when $V\to0$~\cite{Gol13}.

According to Eq.~\eqref{sumphase}, the phase-dependent terms in Eq.~\eqref{ptot} become
\begin{eqnarray}\label{pcosdecomposed}\nonumber
\cos\varphi \Pc&=&\cos(\phi+\sin^{-1}i_b)\Pc=\\
&=&\left (\sqrt{1-i_b^2}\cos\phi-i_b\sin\phi \right )\Pc
\end{eqnarray}
and
\begin{eqnarray}\label{psindecomposed}\nonumber
\sin\varphi \Ps&=&\sin(\phi+\sin^{-1}i_b)\Ps=\\
&=&\left (\sqrt{1-i_b^2}\sin\phi+i_b\cos\phi \right )\Ps.
\end{eqnarray}
Accordingly, $P_{\text{tot}}$ can be recast by defining two phase-dependent terms
\begin{equation}\label{sumphase2}
P_{\text{tot}}(T_1,T_2,\varphi)=\Pqp(T_1,T_2,\varphi)+P_{\phi}^0(T_1,T_2)+P_{\phi}^1(T_1,T_2),
\end{equation}
which depends on the bias current according to
\begin{eqnarray}\label{panomalous1}
P_{\phi}^0(T_1,T_2)&\text{=}&\left (-\cos\phi \Pc+\sin\phi \Ps \right )\sqrt{1-i_b^2}\qquad\\\label{panomalous1}
P_{\phi}^1(T_1,T_2)&\text{=}&\left (\sin\phi \Pc+\cos\phi \Ps \right )i_b.\qquad
\end{eqnarray}

Notably, for $i_b=0$ the usual energy transport across a tunnel junction is recovered 
\begin{eqnarray}\label{sumphase2}\nonumber
P_{\text{tot}}(T_1,T_2,\varphi)&=&\Pqp(T_1,T_2,\varphi)-\cos\phi \Pc(T_1,T_2,\varphi)\\
&&+\sin\phi \Ps(T_1,T_2,\varphi),
\end{eqnarray}
instead, in the limit of $i_b\to1$, Eq.~\eqref{ptot} becomes
\begin{eqnarray}\label{sumphase3}\nonumber
P_{\text{tot}}(T_1,T_2,\varphi)&\to&\Pqp(T_1,T_2,\varphi)+\sin\phi \Pc(T_1,T_2,\varphi)\\
&&+\cos\phi \Ps(T_1,T_2,\varphi).
\end{eqnarray}
Interestingly, by increasing the bias current the role of $\sin\phi$ and $\cos\phi$ tends to swap.
Moreover, we observe that by inverting the flowing direction of the bias current only the sign of $P_{\phi}^1$ changes, see Eq.~\eqref{panomalous1}, whereas $P_{\phi}^0$ is invariant with respect to changes of sign of the bias current.

\begin{figure}[t]
\includegraphics[width=\columnwidth]{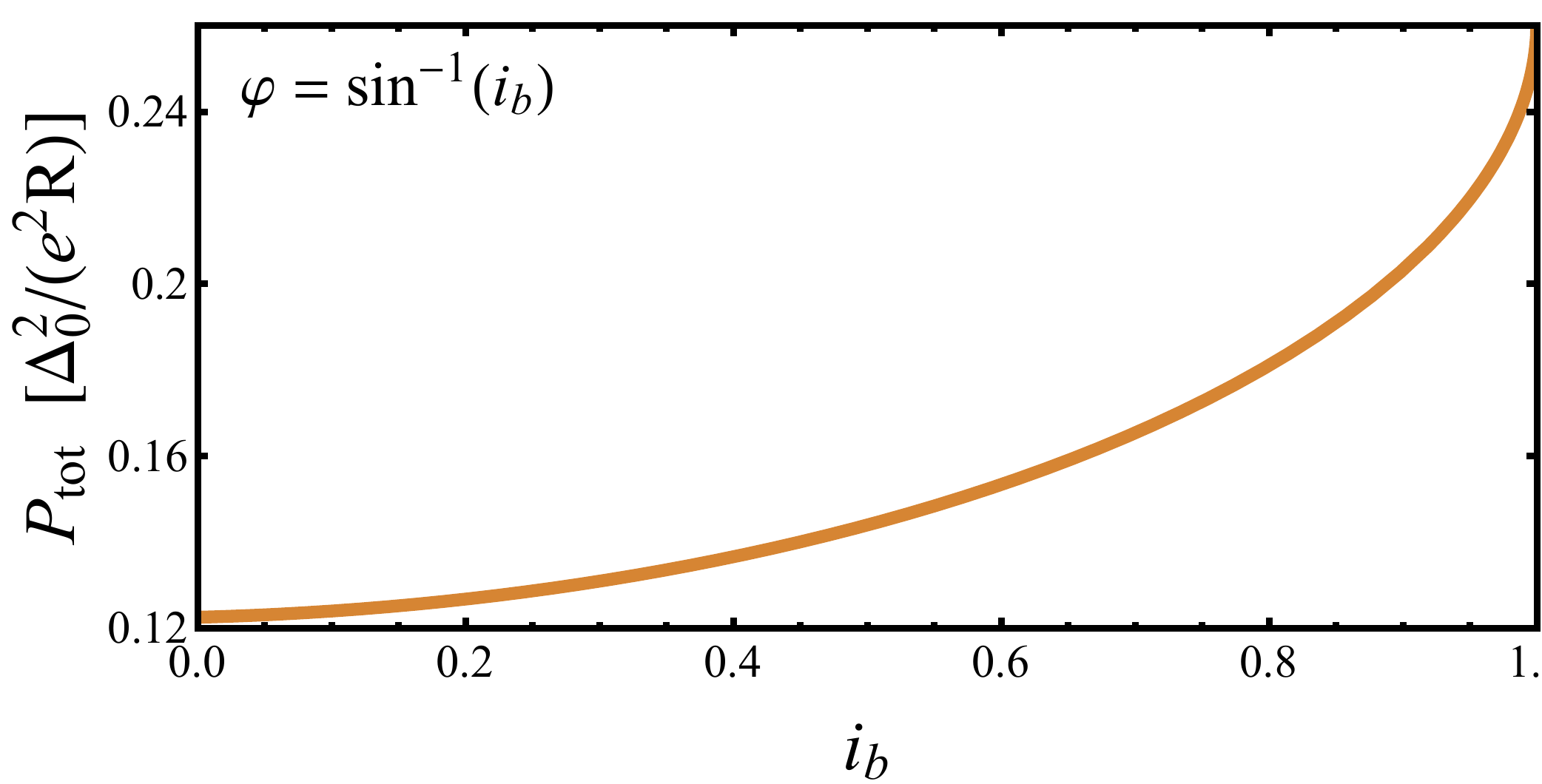}
\caption{$P_{\text{tot}}$ (in units of $\Delta_0^2/(e^2R)$) as a function of the normalized bias current $i_b$, in the absence of soliton, namely, $\phi=0$ in Eq.~\eqref{sumphase} so that $\varphi=\sin^{-1}(i_b)$, for $T_1=7\K$ and $T_2=4.2\K$.}
\label{Fig02}
\end{figure}

The behavior of $P_{\text{tot}}$ in the absence of solitons, namely, $\phi=0$ in Eq.~\eqref{sumphase} so that $\varphi=\sin^{-1}(i_b)$, as a function of the normalized bias current $i_b$ is shown in Fig.~\eqref{Fig02}, for $T_1=7\K$ and $T_2=4.2\K$. We observe that the bias current causes the power $P_{\text{tot}}$ to monotonically increases. 

Finally, we note that in the time-independent case an equilibrium dissipationless current does not generate any Joule heating terms contributing to the energy exchange.


%

\section{Soliton dynamics in an electrically biased LJJ}
\label{Sec02}\vskip-0.2cm

Although LJJs were first measured more than 40 years ago~\cite{Sco69,Ful73}, they are still the subject of both theoretical~\cite{Gul07,Mon12,Val14,Zel15,Pan15,GuaValSpa16,GuaSol17,Hil18,Wus18} and experimental~\cite{Ooi07,Lik12,Fed12,Mon13,Gra14,Kos14,Fed14,Vet15,Gol17} studies, also because they are the ideal solid-state environment to investigate the properties of soliton~\cite{Par93,Ust98}. These excitations give rise to step structures in the I-V characteristic of LJJs, microwaves radiation emission, and they carry a quantum of magnetic flux, $\mathbf{\Phi}_0$, induced by a supercurrent loop surrounding it, with the local magnetic field perpendicularly oriented with respect to the junction length~\cite{McL78}. For this reason, solitons in the context of LJJs are usually referred to as fluxons or Josephson vortices. Solitons in LJJs can be easily generated by an external magnetic field~\cite{Gua18,GuaSolBra18}. Alternatively, in an annular geometry~\cite{Dav86}, i.e., a ``closed'' LJJ folded back into itself in which solitons move with no interactions with edges, fluxons can be excited at will~\cite{Ust92,Ust02}, permitting highly-controllable soliton dynamics.

We consider a current biased long Josephson tunnel junction, with normal-state resistance $R$ and specific capacitance $C_s$, connecting two superconducting leads, $S_1$ and $S_2$ residing at temperatures $T_1$ and $T_2$, see Fig.~\ref{Fig01}(a). We assume leads made by the same superconductor, so that $T_{c_1}=T_{c_2}=T_c$ and $\Delta_0=1.764k_BT_c$ is the zero-temperature superconducting gap.

A bias current flowing through the junction acts on the soliton with a Lorentz force, $\mathbf{F}_L\propto\mathbf{I}_\text{bias}\times\mathbf{\Phi}_0$ [with the direction of $\mathbf{\Phi}_0$ depending on the polarity of the soliton, see Eq.~\eqref{soliton}], see Fig.~\ref{Fig01}(b). So, in the presence of an external bias current a soliton shifts along the junction.

The phase solution representing a soliton moving with velocity $u$ along a LJJ, in the presence of a bias current $I_{\text{bias}}$, can be written approximatively as~\cite{PedSam84}
\begin{equation}\label{soliton}
\varphi_s(x,t)\simeq4\arctan\Big\{ \exp \Big[ \sigma \xi (x,t) \Big]\Big\} +\sin^{-1}\left (i_b \right ),
\end{equation}
where
\begin{equation}\label{csi}
\xi (x,t)=\frac{x-x_0-ut}{\lambda _J\sqrt{1-\left ( \frac{u}{\bar{c}} \right )^2}}=\frac{\widetilde{x}-\widetilde{x}_0-\widetilde{u}\widetilde{t} }{\sqrt{1-\widetilde{u}^2}}
\end{equation}
and $\sigma=\pm1$ is the polarity of the soliton. 
Here, we used the normalized units $\widetilde{x}=x/\lambda_J$, $\widetilde{t}=\omega_pt$, and $\widetilde{u}=u/\bar{c}$, with $\omega_p=\sqrt{\frac{2\pi}{\Phi_0}\frac{J_c}{C_s}}$ and $\lambda_{_{J}}=\sqrt{\frac{\Phi_0}{2\pi \mu_0}\frac{1}{t_d J_c}}$ being the plasma frequency and the Josephson penetration depth, respectively. Moreover, $\mu_0$ is the vacuum permeability, $t_d=\lambda_{L,1}+\lambda_{L,2}+d$ is the effective magnetic thickness (where $\lambda_{L,i}$ is the London penetration depth of the $i$-th superconductor and $d$ is the insulating thickness), and $\bar{c}=\omega_p\lambda_J$ is the Swihart velocity, namely, the limiting soliton velocity in the junction~\cite{Bar82}. The Swihart velocity of typical high-quality superconductor-insulator-superconductor (SIS) junctions is $\bar{c}\sim10^6-10^7\;\text{m}/\text{s}$. 
The velocity-dependent factor in Eq.~\eqref{csi} represents the relativistic contraction of the soliton when its velocity approaches the maximum speed, i.e., $\widetilde{u}\to1$~\cite{McL78}. This is the consequence of Lorentz invariance of the unperturbed sine-Gordon equation describing the electrodynamics of a LJJ~\cite{Bar82}.
The width of the soliton is 
\begin{equation}\label{solitonwidth}
W_s=\lambda _J\sqrt{1-\left ( \frac{u}{\bar{c}} \right )^2}, 
\end{equation}
so that the faster the soliton, the narrower it is.

\begin{figure}[t!!]
\includegraphics[width=\columnwidth]{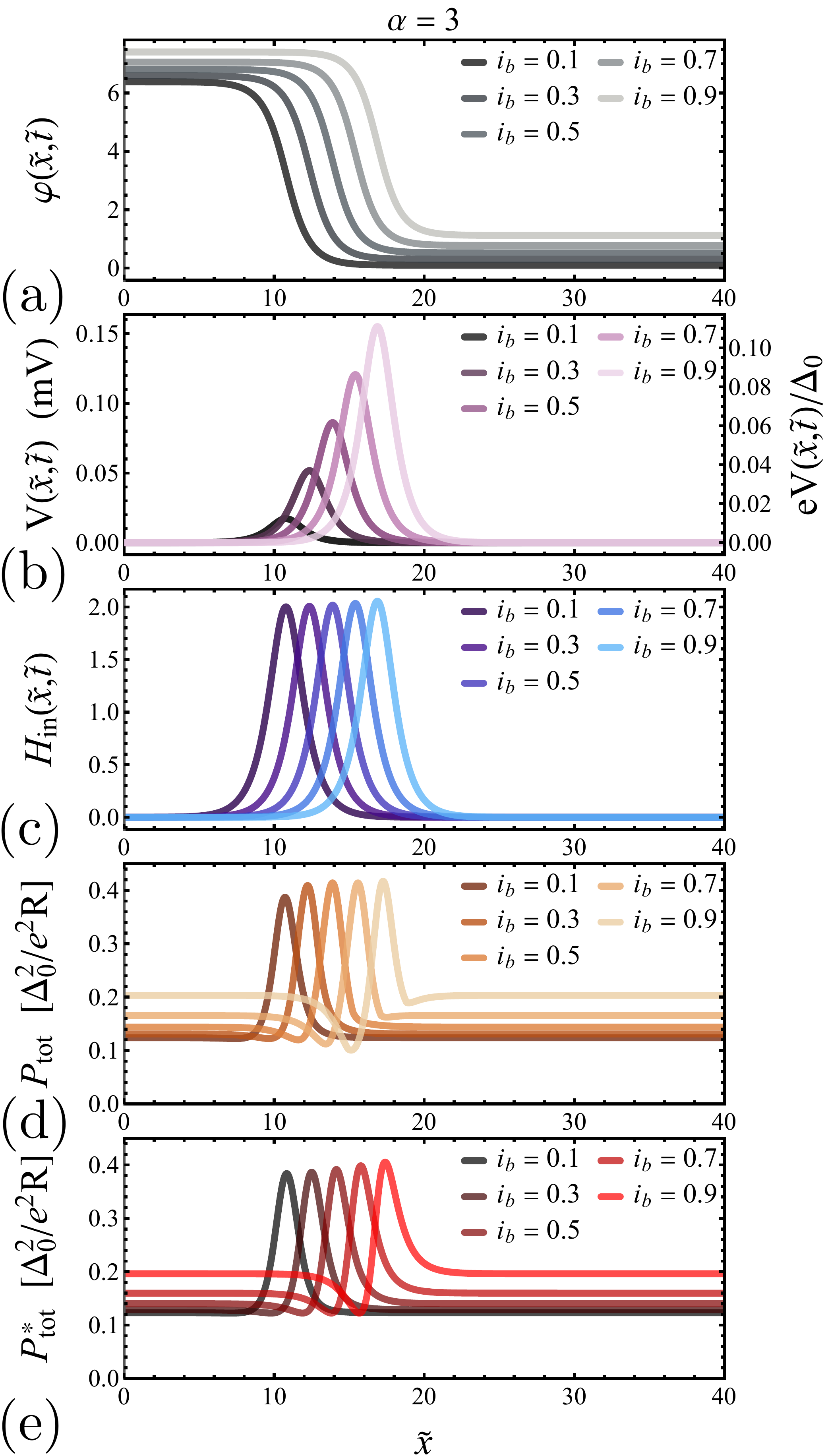}
\caption{Phase evolution (a), voltage drop (b), space derivative of the phase (c), energy transport (d), and heat transport (e) for a soliton moving in a LJJ for $\alpha=3$, at a few values of the bias current $i_b$. The other parameters are: $L=40$, $\widetilde{x}_0=L/4$, $\widetilde{t}=30$, $\omega_p=1\;\text{THz}$, $T_1=7\K$ and $T_2=4.2\K$. The heat currents are in units of $\Delta_0^2/(e^2R)$.}
\label{Fig03}
\end{figure}
\begin{figure}[t!!]
\includegraphics[width=\columnwidth]{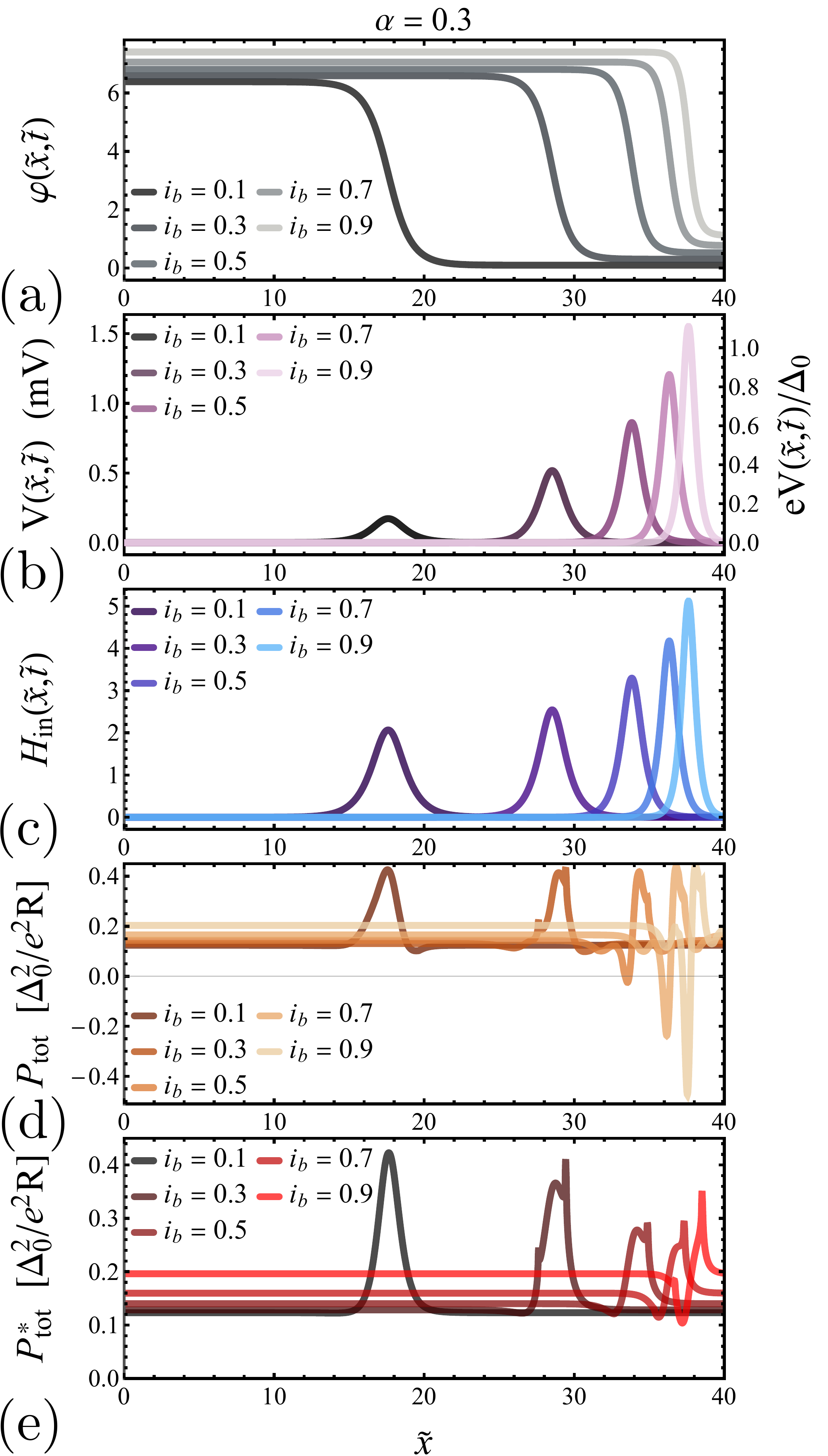}
\caption{Phase evolution (a), voltage drop (b), space derivative of the phase (c), energy transport (d), and heat transport (e) for a soliton moving in a LJJ for $\alpha=0.3$, at a few values of the bias current $i_b$. The other parameters are: $L=40$, $\widetilde{x}_0=L/4$, $\widetilde{t}=30$, $\omega_p=1\;\text{THz}$, $T_1=7\K$ and $T_2=4.2\K$. The heat currents are in units of $\Delta_0^2/(e^2R)$.}
\label{Fig04}
\end{figure}

According to the perturbational approach~\cite{McL78}, in the presence of an external bias current the steady-state drifting velocity of the soliton, in units of $\bar{c}$, reads~\cite{McL78,Bar82}
\begin{equation}\label{driftvelocity}
\widetilde{u}_{d}(i_b)= \frac{1}{\sqrt{1+\left (\frac{4\alpha}{\pi i_b} \right )^2}},
\end{equation}
where $\alpha=1/(\omega_pRC_s)$ is the damping parameter, namely, the parameter quantifying the dissipation in the system~\cite{Bar82}. This is the velocity at which the power input from the bias current is equal to the power loss due to damping affecting the soliton dynamics~\cite{McL78}.

A moving soliton locally induces a voltage drop
\begin{equation}\label{solitonvoltage}
V(x,t)=\frac{\hbar}{2e}\frac{\partial \varphi_s}{\partial t}=\frac{\hbar}{e}\frac{\,\text{sech} \left[\xi (x,t)\right] }{\sqrt{1-\widetilde{u}^2}}\widetilde{u}\omega_p
\end{equation}
and generates a magnetic field (in units of $\frac{2\pi\mu_0t_d}{\Phi_0}$)
\begin{equation}\label{solitonmagneticfield}
H_{\text{in}}(x,t)=\frac{\partial \varphi_s}{\partial x}=\frac{2\,\text{sech} \left[\xi (x,t)\right] }{\lambda_J\sqrt{1-\widetilde{u}^2}}.
\end{equation}

From Eq.~\eqref{driftvelocity} one obtains
$\frac{\widetilde{u}_{d}}{\sqrt{1-\widetilde{u}_{d}^2}}=\frac{\pi i_b}{4\alpha}$,
so that Eq.~\eqref{solitonvoltage} for a steadily drifting soliton, i.e., with $\widetilde{u}=\widetilde{u}_{d}$, becomes
\begin{equation}\label{solitonvoltage_2}
V_d(x,t)=\frac{\hbar}{e}\text{sech} \left[\xi (x,t)\right]\frac{\pi i_b}{4\alpha}\omega_p.
\end{equation}

In order to estimate the modifications to energy and thermal transport, we will analyze the profile of the exchanged power along the junction as a function of various parameters. In particular, we will investigate the steady dynamics of the soliton profile under a constant bias current.

\section{Thermal transport}
\label{Sec03}\vskip-0.2cm

In this paragraph we wish to investigate the consequences on energy and thermal transport across the junction in the presence of a steadily drifting soliton under the effect of a current biasing. 

We expect that the energy flowing through the system will produce evidences, such as a potential modification of the temperature of the junction. Anyway, exploring the thermal behavior of the junction one has to distinguish between dissipative and reactive contributions in Eq.~\eqref{ptot}. In fact, with the aim to determine the temperature profile, the total thermal power density to take into account has to contain only the dissipative contributions, namely,
\begin{equation}\label{Pt}
P^*_{\text{tot}}( T_1,T_2,\varphi)=\Pqp( T_1,T_2)-\cos\varphi \;\Pc( T_1,T_2,V),
\end{equation}
since the term $\Ps$ is purely reactive~\cite{Gol13,Vir17}.

Furthermore, the time evolution of the temperatures can be obtained by solving self-consistently both the sine-Gordon equation for the phase dynamics~\cite{Bar82} and the heat balance equation for each electrode~\cite{GuaSolBra18}. Conversely, in the following, in the place of solving numerically these equations, we will exploit the simple closed expressions of both the solitonic phase solution and the stationary speed of a soliton, see Eqs.\eqref{soliton} and~\eqref{driftvelocity}, respectively, to directly to gain insight on energy and thermal transport across the system in the adiabatic regime, see Sec.~\ref{Sec04a}. The solution obtained in this manner well approximate the full solution, since the two equations governing the evolution are weakly coupled.
Finally, we observe that the characteristic timescales of thermal processes may differ from the typical evolution timescale of solitons, and strongly depend on the specific characteristics of the junction, see Sec.~\ref{Sec04b}.

The investigation of the direct effect on the temperature profile is beyond the scope of this work, and we have motivated reason to think that focusing on single-soliton effects is not the appropriate manner to observe experimentally the reported phenomenology. Anyway, for didactic purposes we discuss the single soliton case as the key element for more complex dynamics. 

So, in the next section, we are going to discuss the energy and thermal transport profiles at fixed times as a function the bias current, in both high and low damping case.

\subsection{Results}
\label{Sec03a}\vskip-0.2cm

In this section we discuss the impact of a bias current on the power flowing through a temperature biased LJJ, as a soliton is set in. With the aim to only explore how energy and heat transport is affected by $i_b$, we assume we can work in the adiabatic limit~\cite{Gol13}, and we use Eq.~\eqref{ptot} to calculate the heat current flowing through the junction. The range of validity of the adiabatic limit approximation will be thoroughly discussed in Sec.~\ref{Sec04a}.

We consider a soliton defined by Eq.~\eqref{soliton} which moves with a steady velocity $\widetilde{u}_{d}(i_b)$, see Eq.~\eqref{driftvelocity}, along a junction with length (in units of $\lambda_J$) $L=40$. Specifically, we investigate thermal transport in the presence of a soliton started from the point $\widetilde{x}_0=L/4$ and travelling along the junction for a time $\widetilde{t}=30$. Clearly, the higher the bias current, the faster the soliton and then the farther it arrives in the time $\widetilde{t}$. 
We analyze the heat transport as a function of the position along the junction, for two values of the damping parameter, namely, $\alpha=3$ and $\alpha=0.3$, at a few values of the bias current, see Figs.~\ref{Fig03} and~\ref{Fig04}, respectively. Hereafter, we set the values $\omega_p=1\;\text{THz}$, $T_1=7\K$, $T_2=4.2\K$, and $T_{c}=9.2\K$ (i.e., a Nb-based junction). 

In the following, we will first discuss the high damping case, since in this regime we can safely use the adiabatic approximation~\cite{Gol13} to study the transport across the JJ, and than we make a comparison with the low damping case. 

{\it High damping case. -- } In Fig.~\ref{Fig03} we show the phase profile [panels (a)], the voltage drop [panels (b)], the local magnetic field [panels (c)], the energy transport [panels (d)], and the heat transport [panels (e)] (both in units of $\Delta_0^2/(e^2R)$) in the high damping case. 

The solitonic phase evolutions, at different values of $i_b$, are shown in Fig.~\ref{Fig03}(a). By increasing the bias current, the soliton, namely, the $2\pi$-step in the phase, moves faster, so that at a fixed time $\widetilde{t}$ it moves rightwards, and becomes sharper. In Fig.~\ref{Fig03}(b), we show the corresponding voltage drop distributions by varying $i_b$. In the right vertical axes of this panel the normalized voltage values, $eV/\Delta_0$, are shown. The voltage drop along the junction is peaked in the center of the soliton, see Eq.~\eqref{solitonvoltage}. Furthermore, by increasing the bias current, we observe the voltage peak to become higher and narrower, since the soliton speeds up and shrinks.
In Fig.~\ref{Fig03}(c) we show the local magnetic field, $H_{\text{in}}(x,t)=\partial \varphi/\partial x$, which instead keep roughly the same amplitude in spite of the width is slightly changed by increasing $i_b$, since $\widetilde{u}\ll1$, see Eq.\eqref{solitonmagneticfield}.

The energy and the heat transport, $P_{\text{tot}}$ and $P^*_{\text{tot}}$, see Eqs.~\eqref{ptot} and~\eqref{Pt}, respectively, for $T_1=7\K$ and $T_2=4.2\K$ are shown in Figs.~\ref{Fig03}(d) and (e).
By increasing the bias current, we expect the phase dependence of the energy exchanged $P_{\text{tot}}$ to change its profile shape. Specifically, for $i_b$ close to zero one obtains $V\to0$ according to Eq.~\eqref{solitonvoltage_2}, so that the term $\Ps$ is vanishingly small~\cite{Gol13}. In this case, the $-\cos\phi$ term dominates $P_{\text{tot}}$, see Eq.~\eqref{sumphase2}, which is positive and single peaked in correspondence of the soliton. We essentially already investigate the temperature evolution in this regime in Ref.~\cite{Gua18}.
Conversely, by increasing $i_b$, both the $\Pc$ and $\Ps$ contributions are affected by the average phase shifting, $\sin^{-1}\left (i_b \right )$, and by the generation of a voltage $V$. As a result, the sine-dependence of $P_{\text{tot}}$ tends to emerge. 
Finally, the reactive contribute $\Ps$ is quite small in this case, we observe that the heat power $P^*_{\text{tot}}$ does not differ too much from $P_{\text{tot}}$ (see Figs.~\ref{Fig03}(d)).
Finally we observe that, in the high damping case, see Fig.~\ref{Fig03}(e) for $\alpha=3$, the deformation of $P^*_{\text{tot}}$ may induce a local heating of the electrode $S_2$ (and concurrently a local cooling of $S_1$, see Fig.~\ref{Fig01}(a)), which depends on $i_b$. We expect to see in this regime a temperature profile different than that one in the case previously reported~\cite{Gua18}.

{\it Low damping case. -- } The scenario changes by reducing the damping parameter, see Fig.~\ref{Fig04} for $\alpha=0.3$. In fact, by reducing $\alpha$, the velocity of the soliton, for a given bias current, increases, as well as the distance it covers in the time $\widetilde{t}$. This is why the curves shown in Fig.~\ref{Fig04} tend to overlap less than those in Fig.~\ref{Fig03}. Moreover, the lower $\alpha$, the higher the maximum value of the voltage drop, according to Eq.~\eqref{solitonvoltage_2}, see Fig.~\ref{Fig04}(b). We observe also the substantial contraction of the soliton by increasing $i_b$, see Fig.~\ref{Fig04}(c), which results also a stronger increase of the intensity in the magnetic field peak.
Concerning the energy exchange, we observe that at high values of the voltage drop, the term $\Ps$ in Eq.~\eqref{ptot} can become more effective with respect to the terms $\Pqp$ and $\Pc$, as it has been discussed in Ref.~\onlinecite{Gol13}. This behavior becomes stronger for low $\alpha$ values, since in this case the soliton speed, as well as the local voltage drop, is higher. In fact, we observe that the total power, $P_{\text{tot}}$, flowing through the system behaves quite differently by reducing $\alpha$, see Fig.~\ref{Fig04}(d) for $\alpha=0.3$.
Firstly, the profile of $P_{\text{tot}}$ is single peaked for low bias currents, but it is distorted when $i_b$ is increased. For such a small $\alpha$, if $i_b\to1$ the voltage $V$ significantly enhances and $\Ps$ becomes greater than $\Pc$, so that the $+\cos\phi$ term in Eq.~\eqref{sumphase3} dominates $P_{\text{tot}}$. In this case, we observe a negative peak of $P_{\text{tot}}$, see Fig.~\ref{Fig04}(d), so that the soliton could even induce a localized change of sign in the total exchange of energy between $S_1$ and $S_2$. Notably, the intensity of this peak can be intensified by reducing the damping, since $\widetilde{u}_{d}\to1$ only if $\alpha\to0$. 
Unfortunately, this negative peak is mainly due to the reactive contribution $\Ps$, so it affects less the dissipative heat power $P^*_{\text{tot}}$, which appears however highly distorted when the bias current increases, see Fig.~\ref{Fig04}(e).
Finally, we observe that the peaks shown in Figs.~\ref{Fig04}(d)-(e) stem from the alignment of the singularities of the BCS DOSs in the superconductors~\cite{Har74,Bar82,Gol13,For16,Gua18bis}.

Beyond these well-localized thermal effects induced by the soliton, we observe that the background value of $P^*_{\text{tot}}$, namely, the heat current flowing far from the soliton, tends to increases with $i_b$, see Figs.~\ref{Fig04}(d) and~\ref{Fig03}(d). This means that the mean temperature of the electrode will globally enhances by increasing the bias current, according to what we discussed in Fig.~\ref{Fig02}. 
This behaviour can be understood by considering that an increase of $i_b$ causes an overall slight increase of the phase, see Figs.~\ref{Fig03}(a) and~\ref{Fig04}(a). This means that, by increasing $i_b$, the contribution of the term $\cos\varphi \Pc$, which opposes $\Pqp$ in Eq.~\eqref{ptot}, tends to reduce, resulting in an increases of $P^*_{\text{tot}}$. So the previously discussed modulation of the $P^*_{\text{tot}}$ profile can be eventually detected as a temperature smaller than the average temperature.

As discussed so far, the distortion of $P_{\text{tot}}$ induced by the bias current is stronger for low values of the damping parameter, c.f., Figs.~\ref{Fig03}(d) and~\ref{Fig04}(d), since this case gives faster solitons, higher voltage drops, and therefore an increase of the effectiveness of the $\Ps$ term. Anyway, the reliability of our argument is based on working in the adiabatic regime~\cite{Gol13}. Therefore, the results discussed so far for high bias currents and low damping should be taken with a grain of salt. In the next section, we will discuss the range of validity of the adiabatic limit as a function of the bias current. 

\begin{figure*}[t!!]
\includegraphics[width=\textwidth]{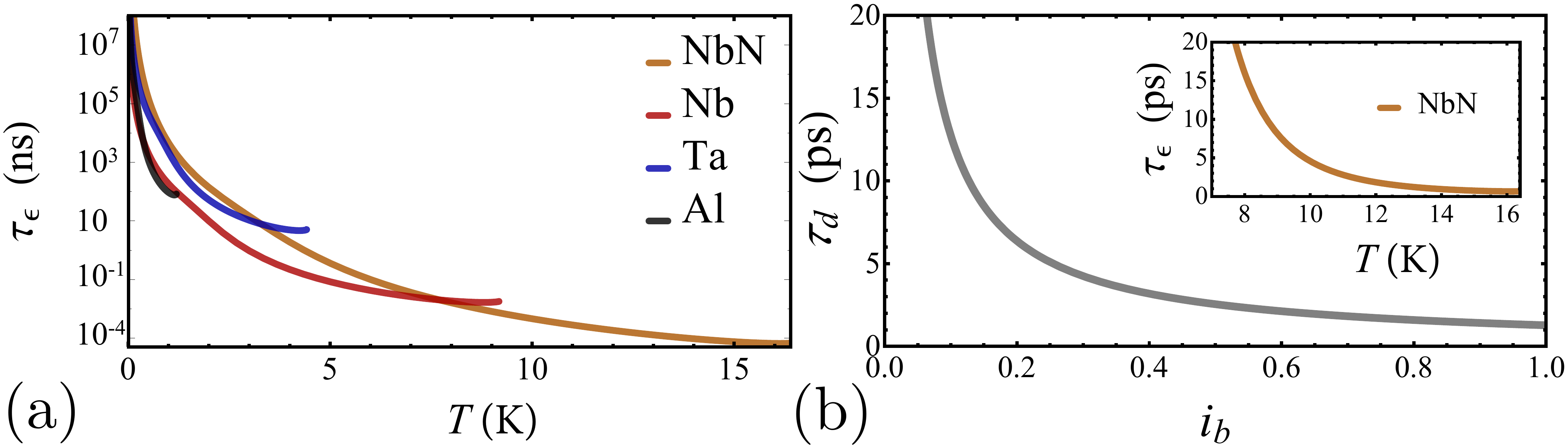}
\caption{(a) Quasiparticle relaxation time, $\tau_\epsilon$, as a function of the temperature calculated for several superconductors, specifically, Al ($T_c=1.2\K$ and $\tau_0=687\ns$~\cite{Bar08}), Ta ($T_c=4.43\K$ and $\tau_0=49\ns$~\cite{Bar08}), Nb ($T_c=9.2\K$ and $\tau_0=0.149\ns$~\cite{Kap76}), and NbN ($T_c=16.4\K$ and $\tau_0=0.5T[\text{K}]^{-1.6}\ns$~\cite{Luo14}). (b) Time $\tau_d$ that a drifting soliton needs to span a length $W_s$ as a function of the bias current, for $\alpha=1$ and $\omega_p=1\;\text{THz}$. In the inset: $\tau_\epsilon$ as a function of the temperature for an NbN junction.}
\label{Fig05}
\end{figure*}

\section{Validity regimes}
\label{Sec04}\vskip-0.2cm

In the previous section we have discussed how the heat transport is affected by a bias driven soliton. Those calculation are based on the validity of the adiabatic approximation. Hereafter, we wish to discuss how in a real system this regime can be safely realized. Finally, we will discuss how the thermal relaxation timescales imply a careful material selection in order to hopefully observe some consequence of the discussed phenomenology.

\subsection{The adiabatic regime}
\label{Sec04a}\vskip-0.2cm

Here, we estimate the range of bias current values according to which the adiabatic regime, and therefore the heat current formulation given by Golubev \emph{et al.}~\cite{Gol13}, holds, in the case of a soliton drifting in a LJJ. 

The adiabatic regime persists for 
\begin{equation}\label{adiabaticregime}
eV\ll \text{Min} \left \{ k_BT_1, k_BT_2, \Delta_1(T_1), \Delta_2(T_2) \right \}.
\end{equation}
Hereafter, for simplicity, we assume $T_1\sim T_2\sim T$ and $\Delta_1\sim\Delta_2\sim\Delta$, since we wish to only give a simple estimation.
In particular, we will compute the upper threshold bias current, $i_{b,\text{th}}$, well below which the adiabaticity is safely satisfied, i.e., $i_b\ll i_{b,\text{th}}$. Indeed, we expect that increasing $i_b$ the soliton moves faster, enhancing the voltage drop, bringing the system out from the adiabatic condition Eq.~\eqref{adiabaticregime}. We take into account the maximum voltage $V_{\text{max}}$ along the junction, namely, the voltage drop in the center of the soliton, $x_{\text{m}}=x_0+vt$, see Figs.~\ref{Fig03}(b) and~\ref{Fig04}(b). 
Since $\text{sech} \left[\xi (x_{\text{m}},t)\right]=1$, for a drifting soliton $e V_{d,\text{max}}$ reads
\begin{equation}\label{solitonvoltagemax}
e V_{d,\text{max}}=\hbar\frac{\pi i_b}{4\alpha}\omega_p.
\end{equation}
Therefore, in order to satisfy Eq.~\eqref{adiabaticregime}, for $T\leq T^*$, with $T^*$ being the temperature at which $k_bT^*=\Delta(T^*)$, one can estimate $i_{b,\text{th}}(T)$ from the relation $e V_{d,\text{max}}=k_BT$, so that 
\begin{equation}\label{ibthreshold}
i_{b,\text{th}}(T)=\frac{4k_B\alpha}{\pi\hbar\omega_p}T=\eta_\alpha T\qquad \text{(for }T\leq T^*\text{)}.
\end{equation}
Instead, for $T> T^*$, $i_{b,\text{th}}(T)$ goes to zero according to $e V_{d,\text{max}}=\Delta(T)$.

For $\alpha=1$, $\omega_p=1\;\text{THz}$, and $T_{c_1}=T_{c_2}=9.2\K$, namely a Nb-based junction, we obtain $T^*\simeq8.31\K$ and $\eta_\alpha\simeq0.17\;\text{K}^{-1}$, so that, for instance, $i_{b,\text{th}}\simeq0.71$ at $T=4.2\K$. 

In conclusion, at a given temperature, the adiabatic condition is satisfied if $i_b\ll i_{b,\text{th}}(T)$, in which case we can advisedly use the usual Golubev's formulation~\cite{Gol13} to calculate the heat current flowing through the junction. We observe that, at a fixed temperature, the range of $i_b$ values assuring the adiabatic regime can be enlarged by increasing $\eta_\alpha$, namely, by increasing the damping parameter, since the soliton slows down and the voltage accordingly reduces, and/or by decreasing the plasma frequency, since the voltage reduces according to Eq.~\eqref{solitonvoltage}.

\subsection{Characteristics timescales}
\label{Sec04b}\vskip-0.2cm

To eventually measure a localized heating induced by a soliton moving along the junction, the system needs ``enough'' time to locally adjust its temperature. In fact, although the soliton locally affects the thermal flux, the superconductor takes some time to thermally respond. 
Then, we can minimally assume that, to effectively induce a local temperature variation, the soliton dynamics should be slower than the timescales dictated by the thermalization processes in the system. The characteristic thermalization time can be estimated as the quasiparticle relaxation time $\tau_\epsilon$ in the superconductor, which is given by~\cite{Kap76}
\begin{equation}\label{quasiparticletime}
\tau_\epsilon^{-1}=\tau_s^{-1}+\tau_r^{-1}.
\end{equation}
In the above equation $\tau_s$ and $\tau_r$ represent, respectively, the scattering and recombination lifetimes, defined according to the well known model of quasiparticle energy relaxation developed by Kaplan \emph{et al.}~\cite{Kap76}. The time $\tau_s$ concerns scattering processes involving emission and absorption of a phonon, whereas $\tau_r$ is related to the recombination of two quasiparticles to form a pair, with the emission of a phonon~\cite{Kap76}. 

In Fig.~\ref{Fig05}(a), the quasiparticle relaxation time as a function of the temperature for several superconductors, specifically, Al ($T_c=1.2\K$ and $\tau_0=687\ns$~\cite{Bar08}), Ta ($T_c=4.43\K$ and $\tau_0=49\ns$~\cite{Bar08}), Nb ($T_c=9.2\K$ and $\tau_0=0.149\ns$~\cite{Kap76}), and NbN ($T_c=16.4\K$ and $\tau_0=0.5T[\text{K}]^{-1.6}\ns$~\cite{Luo14}), is shown. As expected, the quasiparticle relaxation time reduces by increasing the temperature and strongly depends on the material. Specifically, the higher the critical temperature, the lower the $\tau_\epsilon$ values that can be achieved. Indeed, the thermalization time of a NbN junction is of the order of $\text{ps}$ at high temperatures.

Finally, in order to estimate the soliton visibility in a temperature profile, we can assume that the temperature is locally affected by a moving soliton, if the latter stays in the same place for a time long enough to permit the temperature of the superconductor to locally adjust. Therefore, the reported phenomenology could be appreciable in temperature if the time $\tau_d$ that the drifting soliton needs to span the characteristic soliton width $W_s$, see Eq.~\eqref{solitonwidth}, is higher than $\tau_\epsilon$, namely, $\tau_d\gtrsim\tau_\epsilon$. The time $\tau_d$ can be estimated as
\begin{equation}
\tau_d=\frac{W_s}{u_d}=\frac{\lambda_J}{\bar{c}}\frac{\sqrt{1-\widetilde{u}_{d}^2}}{\widetilde{u}_{d}}=\frac{1}{\omega_p}\frac{4\alpha}{\pi i_b}.
\end{equation}

In Fig.~\ref{Fig05}(b), the behavior of $\tau_d$ as a function of the bias current is shown, for $\alpha=1$ and $\omega_p=1\;\text{THz}$. We observe that the condition $\tau_d\gtrsim\tau_\epsilon$ can be effectively fulfilled for a NbN junction, in the limit of low bias currents and high temperatures regimes, see the inset of Fig.~\ref{Fig05}(b). Markedly, by using superconductors with a higher critical temperature, the thermalization time $\tau_\epsilon$ further reduces. Moreover, the time $\tau_d$ linearly grows with both $\alpha$ and the inverse of the plasma frequency $\omega_p^{-1}$. In these cases, the localized temperature change induced by a fast moving soliton could be potentially observed.

Finally, with the aim to effectively observe a soliton-induced temperature variations, we suggest that it is convenient to investigate the temperature of the hot electrode of the junction, in order to increase the response time of the superconductor according to Fig.~\ref{Fig05}.

\section{Conclusions}
\label{Sec05}\vskip-0.2cm

In summary, we have investigated theoretically the phase-coherent heat current flowing through a long Josephson tunnel junction in the presence of a soliton driven by a stationary bias current. The latter acts as a force on the soliton, whose steady drift velocity can be written in a closed simple form~\cite{McL78}. 
We analyzed the distribution of heat currents along the junction by varying the bias current and the damping parameter, as a temperature gradient is imposed across the device. We observed that the bias current significantly affects the heat transport; this effect depends on the value of the damping parameter $\alpha$, since the smaller $\alpha$, the faster moves the soliton. 
In fact, although in the case of a slow soliton a localized heating could be observed in the cold electrode, the heat current profile through the junction significantly modifies when the soliton moves with a speed approaching its limit value, namely, the Swihart velocity.

Furthermore, we discussed the range of values of bias current well below which the adiabatic limit~\cite{Gol13} is assured. Here, we observed that for intermediate values of $\alpha$ the discussed phenomenology could produce observable thermal effects. Moreover, we compared the characteristic timescales of thermalization processes for several superconductors and solitonic dynamics, to establish the limiting regimes to eventually appreciate temperature variations locally induced by soliton-sustained thermal transport. Finally, we observe that the adiabatic limit approximation~\cite{Gol13}, and therefore the reliability of the approach developed in this work, could be not so strictly valid in the limits of low damping and high bias current, namely, as the soliton speed significantly grows.

We expect to see stronger effects of the discussed phenomenology for JJs in the flux-flow regime, namely, when solitons are continuously generated by an external magnetic field and shifted by the current along the junction, where the competitive action between moving solitons, their reflection at a border, and their superimposition will increase additively the discussed effects. We deserve that detailed analysis for a future research. 

\begin{acknowledgments}
C.G., A.B., and F.G. acknowledge the European Research Council under the European Union's Seventh Framework Program (FP7/2007-2013)/ERC Grant agreement No.~615187-COMANCHE and the Tuscany Region under the FARFAS 2014 project SCIADRO for partial financial support. 
P.S. and A.B. have received funding from the European Union FP7/2007-2013 under REA Grant agreement No. 630925 -- COHEAT. 
A.B. acknowledges the CNR-CONICET cooperation programme ``Energy conversion in quantum nanoscale hybrid devices'' and the Royal Society though the International Exchanges between the UK and Italy (grant IES R3 170054).
\end{acknowledgments}

\appendix

\section{Thermal Powers.}
\label{AppA}

In the adiabatic regime, the contributes to the energy transport in a temperature-biased JJ read~\cite{Gol13}
\begin{eqnarray}\label{Pqp}\nonumber
&&\Pqp(T_1,T_2,V)=\frac{1}{e^2R}\int_{-\infty}^{\infty} \mathcal{N}_1 ( \varepsilon-eV ,T_1 )\mathcal{N}_2 ( \varepsilon ,T_2 )\\
&&\times(\varepsilon-eV) [ f ( \varepsilon-eV ,T_1 ) -f ( \varepsilon ,T_2 ) ] d\varepsilon,\\ \nonumber\label{Pcos}
&&P_{\text{qp-pair}}( T_1,T_2,V )=-\frac{1}{e^2R}\int_{-\infty}^{\infty} \mathcal{N}_1 ( \varepsilon-eV ,T_1 )\mathcal{N}_2 ( \varepsilon ,T_2 ) \\
&&\times\frac{\Delta_1(T_1)\Delta_2(T_2)}{\varepsilon}[ f ( \varepsilon-eV ,T_1 ) -f ( \varepsilon ,T_2 ) ]d\varepsilon,\label{Pcos}\\\nonumber
&&P_{\text{pair}}(T_1,T_2,V)=\frac{eV}{2\pi e^2R}\iint_{-\infty}^{\infty} d\epsilon_1d\epsilon_2 \frac{\Delta_1(T_1)\Delta_2(T_2)}{E_2}\\\nonumber
&&\times\left [\frac{1-f(E_1,T_1)-f(E_2,T_2)}{\left ( E_1+E_2 \right )^2-e^2V^2}\text{+}\frac{f(E_1,T_1)-f(E_2,T_2)}{\left ( E_1-E_2 \right )^2-e^2V^2}\right ],\\\label{Psin}
\end{eqnarray}
where $E_j=\sqrt{\epsilon_j^2+\Delta_j(T_j)^2}$, $f ( E ,T )=1/\left (1+e^{E/k_BT} \right )$ is the Fermi distribution function, $\mathcal{N}_j\left ( \varepsilon ,T \right )=\left | \text{Re}\left [ \frac{ \varepsilon +i\gamma_j}{\sqrt{(\varepsilon +i\gamma_j) ^2-\Delta _j\left ( T \right )^2}} \right ] \right |$ is the reduced superconducting density of state, with $\Delta_j\left ( T_j \right )$ and $\gamma_j$ being the BCS energy gap and the Dynes broadening parameter~\cite{Dyn78} of the $j$-th electrode, respectively.

These equations derives from processes involving both Cooper pairs and quasiparticles in tunneling through a JJ predicted by Maki and Griffin~\cite{Mak65}. In fact, $\Pqp$ is the heat power density carried by quasiparticle tunneling, namely, it is an incoherent flow of energy through the junction from the hot to the cold electrode~\cite{Mak65,Gia06}. Instead, the ``anomalous'' terms $\Ps$ and $\Pc$ determine the phase-dependent part of the heat transport originating from the energy-carrying tunneling processes involving Cooper pairs and recombination/destruction of Cooper pairs on both sides of the junction.

We note that $\Ps$, in the low current regime is vanishingly small with respect to both $\Pqp$ and $\Pc$ contributions, and it can be, in principle, neglected. In fact, since this term depends linearly on the time derivative of the phase, it could be effective only when the phase rapidly changes. Anyway, we stress that equation~\eqref{Psin} is a purely reactive contributions~\cite{Gol13,Vir17}, so that in writing a thermal balance equation~\cite{GuaSolBra18} we have to properly neglect it. 


\begin{thebibliography}{59}%
\makeatletter
\providecommand \@ifxundefined [1]{%
 \@ifx{#1\undefined}
}%
\providecommand \@ifnum [1]{%
 \ifnum #1\expandafter \@firstoftwo
 \else \expandafter \@secondoftwo
 \fi
}%
\providecommand \@ifx [1]{%
 \ifx #1\expandafter \@firstoftwo
 \else \expandafter \@secondoftwo
 \fi
}%
\providecommand \natexlab [1]{#1}%
\providecommand \enquote  [1]{``#1''}%
\providecommand \bibnamefont  [1]{#1}%
\providecommand \bibfnamefont [1]{#1}%
\providecommand \citenamefont [1]{#1}%
\providecommand \href@noop [0]{\@secondoftwo}%
\providecommand \href [0]{\begingroup \@sanitize@url \@href}%
\providecommand \@href[1]{\@@startlink{#1}\@@href}%
\providecommand \@@href[1]{\endgroup#1\@@endlink}%
\providecommand \@sanitize@url [0]{\catcode `\\12\catcode `\$12\catcode
  `\&12\catcode `\#12\catcode `\^12\catcode `\_12\catcode `\%12\relax}%
\providecommand \@@startlink[1]{}%
\providecommand \@@endlink[0]{}%
\providecommand \url  [0]{\begingroup\@sanitize@url \@url }%
\providecommand \@url [1]{\endgroup\@href {#1}{\urlprefix }}%
\providecommand \urlprefix  [0]{URL }%
\providecommand \Eprint [0]{\href }%
\providecommand \doibase [0]{http://dx.doi.org/}%
\providecommand \selectlanguage [0]{\@gobble}%
\providecommand \bibinfo  [0]{\@secondoftwo}%
\providecommand \bibfield  [0]{\@secondoftwo}%
\providecommand \translation [1]{[#1]}%
\providecommand \BibitemOpen [0]{}%
\providecommand \bibitemStop [0]{}%
\providecommand \bibitemNoStop [0]{.\EOS\space}%
\providecommand \EOS [0]{\spacefactor3000\relax}%
\providecommand \BibitemShut  [1]{\csname bibitem#1\endcsname}%
\let\auto@bib@innerbib\@empty
\bibitem [{\citenamefont {Parmentier}(1993)}]{Par93}%
  \BibitemOpen
  \bibfield  {author} {\bibinfo {author} {\bibfnamefont {R.~D.}\ \bibnamefont
  {Parmentier}},\ }\enquote {\bibinfo {title} {Solitons and long Josephson
  junctions},}\ in\ \href {\doibase 10.1007/978-94-011-1918-4_7} {\emph
  {\bibinfo {booktitle} {The New Superconducting Electronics}}},\ \bibinfo
  {editor} {edited by\ \bibinfo {editor} {\bibfnamefont {H.}~\bibnamefont
  {Weinstock}}\ and\ \bibinfo {editor} {\bibfnamefont {R.~W.}\ \bibnamefont
  {Ralston}}}\ (\bibinfo  {publisher} {Springer Netherlands},\ \bibinfo
  {address} {Dordrecht},\ \bibinfo {year} {1993})\ pp.\ \bibinfo {pages}
  {221--248}\BibitemShut {NoStop}%
\bibitem [{\citenamefont {Ustinov}(1998)}]{Ust98}%
  \BibitemOpen
  \bibfield  {author} {\bibinfo {author} {\bibfnamefont {A.~V.}\ \bibnamefont
  {Ustinov}},\ }\href@noop {} {\bibfield  {journal} {\bibinfo  {journal}
  {Physica D}\ }\textbf {\bibinfo {volume} {123}},\ \bibinfo {pages} {315}
  (\bibinfo {year} {1998})}\BibitemShut {NoStop}%
\bibitem [{\citenamefont {Guarcello}\ \emph
  {et~al.}(2016{\natexlab{a}})\citenamefont {Guarcello}, \citenamefont
  {Giazotto},\ and\ \citenamefont {Solinas}}]{Gua16}%
  \BibitemOpen
  \bibfield  {author} {\bibinfo {author} {\bibfnamefont {C.}~\bibnamefont
  {Guarcello}}, \bibinfo {author} {\bibfnamefont {F.}~\bibnamefont {Giazotto}},
  \ and\ \bibinfo {author} {\bibfnamefont {P.}~\bibnamefont {Solinas}},\ }\href
  {\doibase 10.1103/PhysRevB.94.054522} {\bibfield  {journal} {\bibinfo
  {journal} {Phys. Rev. B}\ }\textbf {\bibinfo {volume} {94}},\ \bibinfo
  {pages} {054522} (\bibinfo {year} {2016}{\natexlab{a}})}\BibitemShut
  {NoStop}%
\bibitem [{\citenamefont {Guarcello}\ \emph
  {et~al.}(2018{\natexlab{a}})\citenamefont {Guarcello}, \citenamefont
  {Solinas}, \citenamefont {Braggio},\ and\ \citenamefont {Giazotto}}]{Gua18}%
  \BibitemOpen
  \bibfield  {author} {\bibinfo {author} {\bibfnamefont {C.}~\bibnamefont
  {Guarcello}}, \bibinfo {author} {\bibfnamefont {P.}~\bibnamefont {Solinas}},
  \bibinfo {author} {\bibfnamefont {A.}~\bibnamefont {Braggio}}, \ and\
  \bibinfo {author} {\bibfnamefont {F.}~\bibnamefont {Giazotto}},\ }\href
  {\doibase 10.1103/PhysRevApplied.9.034014} {\bibfield  {journal} {\bibinfo
  {journal} {Phys. Rev. Applied}\ }\textbf {\bibinfo {volume} {9}},\ \bibinfo
  {pages} {034014} (\bibinfo {year} {2018}{\natexlab{a}})}\BibitemShut
  {NoStop}%
\bibitem [{\citenamefont {Guarcello}\ \emph
  {et~al.}(2018{\natexlab{b}})\citenamefont {Guarcello}, \citenamefont
  {Solinas}, \citenamefont {Braggio},\ and\ \citenamefont
  {Giazotto}}]{GuaSolBra18}%
  \BibitemOpen
  \bibfield  {author} {\bibinfo {author} {\bibfnamefont {C.}~\bibnamefont
  {Guarcello}}, \bibinfo {author} {\bibfnamefont {P.}~\bibnamefont {Solinas}},
  \bibinfo {author} {\bibfnamefont {A.}~\bibnamefont {Braggio}}, \ and\
  \bibinfo {author} {\bibfnamefont {F.}~\bibnamefont {Giazotto}},\ }\href@noop
  {} {\bibfield  {journal} {\bibinfo  {journal} {arXiv preprint
  arXiv:1803.02588}\ } (\bibinfo {year} {2018}{\natexlab{b}})}\BibitemShut
  {NoStop}%
\bibitem [{\citenamefont {Giazotto}\ \emph {et~al.}(2013)\citenamefont
  {Giazotto}, \citenamefont {Mart\'{\i}nez-P\'erez},\ and\ \citenamefont
  {Solinas}}]{Gia13}%
  \BibitemOpen
  \bibfield  {author} {\bibinfo {author} {\bibfnamefont {F.}~\bibnamefont
  {Giazotto}}, \bibinfo {author} {\bibfnamefont {M.~J.}\ \bibnamefont
  {Mart\'{\i}nez-P\'erez}}, \ and\ \bibinfo {author} {\bibfnamefont
  {P.}~\bibnamefont {Solinas}},\ }\href {\doibase 10.1103/PhysRevB.88.094506}
  {\bibfield  {journal} {\bibinfo  {journal} {Phys. Rev. B}\ }\textbf {\bibinfo
  {volume} {88}},\ \bibinfo {pages} {094506} (\bibinfo {year}
  {2013})}\BibitemShut {NoStop}%
\bibitem [{\citenamefont {Mart{\'\i}nez-P{\'e}rez}\ and\ \citenamefont
  {Giazotto}(2014)}]{Mar14}%
  \BibitemOpen
  \bibfield  {author} {\bibinfo {author} {\bibfnamefont {M.~J.}\ \bibnamefont
  {Mart{\'\i}nez-P{\'e}rez}}\ and\ \bibinfo {author} {\bibfnamefont
  {F.}~\bibnamefont {Giazotto}},\ }\href@noop {} {\bibfield  {journal}
  {\bibinfo  {journal} {Nat. Commun.}\ }\textbf {\bibinfo {volume} {5}},\
  \bibinfo {pages} {3579} (\bibinfo {year} {2014})}\BibitemShut {NoStop}%
\bibitem [{\citenamefont {Giazotto}\ and\ \citenamefont
  {Mart{\'\i}nez-P{\'e}rez}(2012{\natexlab{a}})}]{GiaMar12}%
  \BibitemOpen
  \bibfield  {author} {\bibinfo {author} {\bibfnamefont {F.}~\bibnamefont
  {Giazotto}}\ and\ \bibinfo {author} {\bibfnamefont {M.~J.}\ \bibnamefont
  {Mart{\'\i}nez-P{\'e}rez}},\ }\href {\doibase 10.1063/1.4750068} {\bibfield
  {journal} {\bibinfo  {journal} {Appl. Phys. Lett.}\ }\textbf {\bibinfo
  {volume} {101}},\ \bibinfo {pages} {102601} (\bibinfo {year}
  {2012}{\natexlab{a}})}\BibitemShut {NoStop}%
\bibitem [{\citenamefont {Giazotto}\ and\ \citenamefont
  {Mart{\'\i}nez-P{\'e}rez}(2012{\natexlab{b}})}]{Gia12}%
  \BibitemOpen
  \bibfield  {author} {\bibinfo {author} {\bibfnamefont {F.}~\bibnamefont
  {Giazotto}}\ and\ \bibinfo {author} {\bibfnamefont {M.~J.}\ \bibnamefont
  {Mart{\'\i}nez-P{\'e}rez}},\ }\href@noop {} {\bibfield  {journal} {\bibinfo
  {journal} {Nature}\ }\textbf {\bibinfo {volume} {492}},\ \bibinfo {pages}
  {401} (\bibinfo {year} {2012}{\natexlab{b}})}\BibitemShut {NoStop}%
\bibitem [{\citenamefont {Giazotto}\ \emph {et~al.}(2006)\citenamefont
  {Giazotto}, \citenamefont {Heikkil\"a}, \citenamefont {Luukanen},
  \citenamefont {Savin},\ and\ \citenamefont {Pekola}}]{Gia06}%
  \BibitemOpen
  \bibfield  {author} {\bibinfo {author} {\bibfnamefont {F.}~\bibnamefont
  {Giazotto}}, \bibinfo {author} {\bibfnamefont {T.~T.}\ \bibnamefont
  {Heikkil\"a}}, \bibinfo {author} {\bibfnamefont {A.}~\bibnamefont
  {Luukanen}}, \bibinfo {author} {\bibfnamefont {A.~M.}\ \bibnamefont {Savin}},
  \ and\ \bibinfo {author} {\bibfnamefont {J.~P.}\ \bibnamefont {Pekola}},\
  }\href {\doibase 10.1103/RevModPhys.78.217} {\bibfield  {journal} {\bibinfo
  {journal} {Rev. Mod. Phys.}\ }\textbf {\bibinfo {volume} {78}},\ \bibinfo
  {pages} {217} (\bibinfo {year} {2006})}\BibitemShut {NoStop}%
\bibitem [{\citenamefont {Mart{\'i}nez-P{\'e}rez}\ \emph
  {et~al.}(2014)\citenamefont {Mart{\'i}nez-P{\'e}rez}, \citenamefont
  {Solinas},\ and\ \citenamefont {Giazotto}}]{MarSol14}%
  \BibitemOpen
  \bibfield  {author} {\bibinfo {author} {\bibfnamefont {M.~J.}\ \bibnamefont
  {Mart{\'i}nez-P{\'e}rez}}, \bibinfo {author} {\bibfnamefont {P.}~\bibnamefont
  {Solinas}}, \ and\ \bibinfo {author} {\bibfnamefont {F.}~\bibnamefont
  {Giazotto}},\ }\href {\doibase 10.1007/s10909-014-1132-6} {\bibfield
  {journal} {\bibinfo  {journal} {J. Low Temp. Phys.}\ }\textbf {\bibinfo
  {volume} {175}},\ \bibinfo {pages} {813} (\bibinfo {year}
  {2014})}\BibitemShut {NoStop}%
\bibitem [{\citenamefont {Fornieri}\ and\ \citenamefont
  {Giazotto}(2017)}]{ForGia17}%
  \BibitemOpen
  \bibfield  {author} {\bibinfo {author} {\bibfnamefont {A.}~\bibnamefont
  {Fornieri}}\ and\ \bibinfo {author} {\bibfnamefont {F.}~\bibnamefont
  {Giazotto}},\ }\href@noop {} {\bibfield  {journal} {\bibinfo  {journal} {Nat.
  Nanotechnology}\ }\textbf {\bibinfo {volume} {12}},\ \bibinfo {pages} {944}
  (\bibinfo {year} {2017})}\BibitemShut {NoStop}%
\bibitem [{\citenamefont {Mart{\'\i}nez-P{\'e}rez}\ \emph
  {et~al.}(2015)\citenamefont {Mart{\'\i}nez-P{\'e}rez}, \citenamefont
  {Fornieri},\ and\ \citenamefont {Giazotto}}]{Mar15}%
  \BibitemOpen
  \bibfield  {author} {\bibinfo {author} {\bibfnamefont {M.~J.}\ \bibnamefont
  {Mart{\'\i}nez-P{\'e}rez}}, \bibinfo {author} {\bibfnamefont
  {A.}~\bibnamefont {Fornieri}}, \ and\ \bibinfo {author} {\bibfnamefont
  {F.}~\bibnamefont {Giazotto}},\ }\href@noop {} {\bibfield  {journal}
  {\bibinfo  {journal} {Nature Nanotechnology}\ }\textbf {\bibinfo {volume}
  {10}},\ \bibinfo {pages} {303} (\bibinfo {year} {2015})}\BibitemShut
  {NoStop}%
\bibitem [{\citenamefont {Fornieri}\ \emph {et~al.}(2016)\citenamefont
  {Fornieri}, \citenamefont {Timossi}, \citenamefont {Bosisio}, \citenamefont
  {Solinas},\ and\ \citenamefont {Giazotto}}]{For16}%
  \BibitemOpen
  \bibfield  {author} {\bibinfo {author} {\bibfnamefont {A.}~\bibnamefont
  {Fornieri}}, \bibinfo {author} {\bibfnamefont {G.}~\bibnamefont {Timossi}},
  \bibinfo {author} {\bibfnamefont {R.}~\bibnamefont {Bosisio}}, \bibinfo
  {author} {\bibfnamefont {P.}~\bibnamefont {Solinas}}, \ and\ \bibinfo
  {author} {\bibfnamefont {F.}~\bibnamefont {Giazotto}},\ }\href {\doibase
  10.1103/PhysRevB.93.134508} {\bibfield  {journal} {\bibinfo  {journal} {Phys.
  Rev. B}\ }\textbf {\bibinfo {volume} {93}},\ \bibinfo {pages} {134508}
  (\bibinfo {year} {2016})}\BibitemShut {NoStop}%
\bibitem [{\citenamefont {Guarcello}\ \emph
  {et~al.}(2017{\natexlab{a}})\citenamefont {Guarcello}, \citenamefont
  {Solinas}, \citenamefont {Di~Ventra},\ and\ \citenamefont
  {Giazotto}}]{Gua17}%
  \BibitemOpen
  \bibfield  {author} {\bibinfo {author} {\bibfnamefont {C.}~\bibnamefont
  {Guarcello}}, \bibinfo {author} {\bibfnamefont {P.}~\bibnamefont {Solinas}},
  \bibinfo {author} {\bibfnamefont {M.}~\bibnamefont {Di~Ventra}}, \ and\
  \bibinfo {author} {\bibfnamefont {F.}~\bibnamefont {Giazotto}},\ }\href
  {\doibase 10.1103/PhysRevApplied.7.044021} {\bibfield  {journal} {\bibinfo
  {journal} {Phys. Rev. Applied}\ }\textbf {\bibinfo {volume} {7}},\ \bibinfo
  {pages} {044021} (\bibinfo {year} {2017}{\natexlab{a}})}\BibitemShut
  {NoStop}%
\bibitem [{\citenamefont {Guarcello}\ \emph
  {et~al.}(2018{\natexlab{c}})\citenamefont {Guarcello}, \citenamefont
  {Solinas}, \citenamefont {Braggio}, \citenamefont {Di~Ventra},\ and\
  \citenamefont {Giazotto}}]{GuaSol18}%
  \BibitemOpen
  \bibfield  {author} {\bibinfo {author} {\bibfnamefont {C.}~\bibnamefont
  {Guarcello}}, \bibinfo {author} {\bibfnamefont {P.}~\bibnamefont {Solinas}},
  \bibinfo {author} {\bibfnamefont {A.}~\bibnamefont {Braggio}}, \bibinfo
  {author} {\bibfnamefont {M.}~\bibnamefont {Di~Ventra}}, \ and\ \bibinfo
  {author} {\bibfnamefont {F.}~\bibnamefont {Giazotto}},\ }\href {\doibase
  10.1103/PhysRevApplied.9.014021} {\bibfield  {journal} {\bibinfo  {journal}
  {Phys. Rev. Applied}\ }\textbf {\bibinfo {volume} {9}},\ \bibinfo {pages}
  {014021} (\bibinfo {year} {2018}{\natexlab{c}})}\BibitemShut {NoStop}%
\bibitem [{\citenamefont {Solinas}\ \emph {et~al.}(2016)\citenamefont
  {Solinas}, \citenamefont {Bosisio},\ and\ \citenamefont {Giazotto}}]{Sol16}%
  \BibitemOpen
  \bibfield  {author} {\bibinfo {author} {\bibfnamefont {P.}~\bibnamefont
  {Solinas}}, \bibinfo {author} {\bibfnamefont {R.}~\bibnamefont {Bosisio}}, \
  and\ \bibinfo {author} {\bibfnamefont {F.}~\bibnamefont {Giazotto}},\ }\href
  {\doibase 10.1103/PhysRevB.93.224521} {\bibfield  {journal} {\bibinfo
  {journal} {Phys. Rev. B}\ }\textbf {\bibinfo {volume} {93}},\ \bibinfo
  {pages} {224521} (\bibinfo {year} {2016})}\BibitemShut {NoStop}%
\bibitem [{\citenamefont {Paolucci}\ \emph
  {et~al.}(2017{\natexlab{a}})\citenamefont {Paolucci}, \citenamefont
  {Marchegiani}, \citenamefont {Strambini},\ and\ \citenamefont
  {Giazotto}}]{Pao18}%
  \BibitemOpen
  \bibfield  {author} {\bibinfo {author} {\bibfnamefont {F.}~\bibnamefont
  {Paolucci}}, \bibinfo {author} {\bibfnamefont {G.}~\bibnamefont
  {Marchegiani}}, \bibinfo {author} {\bibfnamefont {E.}~\bibnamefont
  {Strambini}}, \ and\ \bibinfo {author} {\bibfnamefont {F.}~\bibnamefont
  {Giazotto}},\ }\href@noop {} {\bibfield  {journal} {\bibinfo  {journal}
  {arXiv preprint arXiv:1709.08609}\ } (\bibinfo {year}
  {2017}{\natexlab{a}})}\BibitemShut {NoStop}%
\bibitem [{\citenamefont {Timossi}\ \emph {et~al.}(2018)\citenamefont
  {Timossi}, \citenamefont {Fornieri}, \citenamefont {Paolucci}, \citenamefont
  {Puglia},\ and\ \citenamefont {Giazotto}}]{Tim18}%
  \BibitemOpen
  \bibfield  {author} {\bibinfo {author} {\bibfnamefont {G.~F.}\ \bibnamefont
  {Timossi}}, \bibinfo {author} {\bibfnamefont {A.}~\bibnamefont {Fornieri}},
  \bibinfo {author} {\bibfnamefont {F.}~\bibnamefont {Paolucci}}, \bibinfo
  {author} {\bibfnamefont {C.}~\bibnamefont {Puglia}}, \ and\ \bibinfo {author}
  {\bibfnamefont {F.}~\bibnamefont {Giazotto}},\ }\href {\doibase
  10.1021/acs.nanolett.7b04906} {\bibfield  {journal} {\bibinfo  {journal}
  {Nano Letters}\ }\textbf {\bibinfo {volume} {18}},\ \bibinfo {pages} {1764}
  (\bibinfo {year} {2018})}\BibitemShut {NoStop}%
\bibitem [{\citenamefont {Paolucci}\ \emph
  {et~al.}(2017{\natexlab{b}})\citenamefont {Paolucci}, \citenamefont
  {Marchegiani}, \citenamefont {Strambini},\ and\ \citenamefont
  {Giazotto}}]{Pao17}%
  \BibitemOpen
  \bibfield  {author} {\bibinfo {author} {\bibfnamefont {F.}~\bibnamefont
  {Paolucci}}, \bibinfo {author} {\bibfnamefont {G.}~\bibnamefont
  {Marchegiani}}, \bibinfo {author} {\bibfnamefont {E.}~\bibnamefont
  {Strambini}}, \ and\ \bibinfo {author} {\bibfnamefont {F.}~\bibnamefont
  {Giazotto}},\ }\href@noop {} {\bibfield  {journal} {\bibinfo  {journal} {EPL
  (Europhysics Letters)}\ }\textbf {\bibinfo {volume} {118}},\ \bibinfo {pages}
  {68004} (\bibinfo {year} {2017}{\natexlab{b}})}\BibitemShut {NoStop}%
\bibitem [{\citenamefont {Barone}\ and\ \citenamefont
  {Patern\`{o}}(1982)}]{Bar82}%
  \BibitemOpen
  \bibfield  {author} {\bibinfo {author} {\bibfnamefont {A.}~\bibnamefont
  {Barone}}\ and\ \bibinfo {author} {\bibfnamefont {G.}~\bibnamefont
  {Patern\`{o}}},\ }\href@noop {} {\emph {\bibinfo {title} {Physics and
  Applications of the Josephson Effect}}}\ (\bibinfo  {publisher} {Wiley, New
  York},\ \bibinfo {year} {1982})\BibitemShut {NoStop}%
\bibitem [{\citenamefont {Pedersen}\ \emph {et~al.}(1984)\citenamefont
  {Pedersen}, \citenamefont {Samuelsen},\ and\ \citenamefont
  {Welner}}]{PedSam84}%
  \BibitemOpen
  \bibfield  {author} {\bibinfo {author} {\bibfnamefont {N.~F.}\ \bibnamefont
  {Pedersen}}, \bibinfo {author} {\bibfnamefont {M.~R.}\ \bibnamefont
  {Samuelsen}}, \ and\ \bibinfo {author} {\bibfnamefont {D.}~\bibnamefont
  {Welner}},\ }\href {\doibase 10.1103/PhysRevB.30.4057} {\bibfield  {journal}
  {\bibinfo  {journal} {Phys. Rev. B}\ }\textbf {\bibinfo {volume} {30}},\
  \bibinfo {pages} {4057} (\bibinfo {year} {1984})}\BibitemShut {NoStop}%
\bibitem [{\citenamefont {McLaughlin}\ and\ \citenamefont
  {Scott}(1978)}]{McL78}%
  \BibitemOpen
  \bibfield  {author} {\bibinfo {author} {\bibfnamefont {D.~W.}\ \bibnamefont
  {McLaughlin}}\ and\ \bibinfo {author} {\bibfnamefont {A.~C.}\ \bibnamefont
  {Scott}},\ }\href {\doibase 10.1103/PhysRevA.18.1652} {\bibfield  {journal}
  {\bibinfo  {journal} {Phys. Rev. A}\ }\textbf {\bibinfo {volume} {18}},\
  \bibinfo {pages} {1652} (\bibinfo {year} {1978})}\BibitemShut {NoStop}%
\bibitem [{\citenamefont {Guttman}\ \emph
  {et~al.}(1997{\natexlab{a}})\citenamefont {Guttman}, \citenamefont
  {Nathanson}, \citenamefont {Ben-Jacob},\ and\ \citenamefont
  {Bergman}}]{Gut97}%
  \BibitemOpen
  \bibfield  {author} {\bibinfo {author} {\bibfnamefont {G.~D.}\ \bibnamefont
  {Guttman}}, \bibinfo {author} {\bibfnamefont {B.}~\bibnamefont {Nathanson}},
  \bibinfo {author} {\bibfnamefont {E.}~\bibnamefont {Ben-Jacob}}, \ and\
  \bibinfo {author} {\bibfnamefont {D.~J.}\ \bibnamefont {Bergman}},\ }\href
  {\doibase 10.1103/PhysRevB.55.3849} {\bibfield  {journal} {\bibinfo
  {journal} {Phys. Rev. B}\ }\textbf {\bibinfo {volume} {55}},\ \bibinfo
  {pages} {3849} (\bibinfo {year} {1997}{\natexlab{a}})}\BibitemShut {NoStop}%
\bibitem [{\citenamefont {Guttman}\ \emph
  {et~al.}(1997{\natexlab{b}})\citenamefont {Guttman}, \citenamefont
  {Nathanson}, \citenamefont {Ben-Jacob},\ and\ \citenamefont
  {Bergman}}]{GutNat97}%
  \BibitemOpen
  \bibfield  {author} {\bibinfo {author} {\bibfnamefont {G.~D.}\ \bibnamefont
  {Guttman}}, \bibinfo {author} {\bibfnamefont {B.}~\bibnamefont {Nathanson}},
  \bibinfo {author} {\bibfnamefont {E.}~\bibnamefont {Ben-Jacob}}, \ and\
  \bibinfo {author} {\bibfnamefont {D.~J.}\ \bibnamefont {Bergman}},\ }\href
  {\doibase 10.1103/PhysRevB.55.12691} {\bibfield  {journal} {\bibinfo
  {journal} {Phys. Rev. B}\ }\textbf {\bibinfo {volume} {55}},\ \bibinfo
  {pages} {12691} (\bibinfo {year} {1997}{\natexlab{b}})}\BibitemShut {NoStop}%
\bibitem [{\citenamefont {Golubev}\ \emph {et~al.}(2013)\citenamefont
  {Golubev}, \citenamefont {Faivre},\ and\ \citenamefont {Pekola}}]{Gol13}%
  \BibitemOpen
  \bibfield  {author} {\bibinfo {author} {\bibfnamefont {D.}~\bibnamefont
  {Golubev}}, \bibinfo {author} {\bibfnamefont {T.}~\bibnamefont {Faivre}}, \
  and\ \bibinfo {author} {\bibfnamefont {J.~P.}\ \bibnamefont {Pekola}},\
  }\href {\doibase 10.1103/PhysRevB.87.094522} {\bibfield  {journal} {\bibinfo
  {journal} {Phys. Rev. B}\ }\textbf {\bibinfo {volume} {87}},\ \bibinfo
  {pages} {094522} (\bibinfo {year} {2013})}\BibitemShut {NoStop}%
\bibitem [{\citenamefont {Virtanen}\ \emph {et~al.}(2017)\citenamefont
  {Virtanen}, \citenamefont {Solinas},\ and\ \citenamefont {Giazotto}}]{Vir17}%
  \BibitemOpen
  \bibfield  {author} {\bibinfo {author} {\bibfnamefont {P.}~\bibnamefont
  {Virtanen}}, \bibinfo {author} {\bibfnamefont {P.}~\bibnamefont {Solinas}}, \
  and\ \bibinfo {author} {\bibfnamefont {F.}~\bibnamefont {Giazotto}},\ }\href
  {\doibase 10.1103/PhysRevB.95.144512} {\bibfield  {journal} {\bibinfo
  {journal} {Phys. Rev. B}\ }\textbf {\bibinfo {volume} {95}},\ \bibinfo
  {pages} {144512} (\bibinfo {year} {2017})}\BibitemShut {NoStop}%
\bibitem [{\citenamefont {Maki}\ and\ \citenamefont {Griffin}(1965)}]{Mak65}%
  \BibitemOpen
  \bibfield  {author} {\bibinfo {author} {\bibfnamefont {K.}~\bibnamefont
  {Maki}}\ and\ \bibinfo {author} {\bibfnamefont {A.}~\bibnamefont {Griffin}},\
  }\href {\doibase 10.1103/PhysRevLett.15.921} {\bibfield  {journal} {\bibinfo
  {journal} {Phys. Rev. Lett.}\ }\textbf {\bibinfo {volume} {15}},\ \bibinfo
  {pages} {921} (\bibinfo {year} {1965})}\BibitemShut {NoStop}%
\bibitem [{\citenamefont {Frank}\ and\ \citenamefont {Krech}(1997)}]{Fra97}%
  \BibitemOpen
  \bibfield  {author} {\bibinfo {author} {\bibfnamefont {B.}~\bibnamefont
  {Frank}}\ and\ \bibinfo {author} {\bibfnamefont {W.}~\bibnamefont {Krech}},\
  }\href {\doibase http://dx.doi.org/10.1016/S0375-9601(97)00627-0} {\bibfield
  {journal} {\bibinfo  {journal} {Phys. Lett. A}\ }\textbf {\bibinfo {volume}
  {235}},\ \bibinfo {pages} {281 } (\bibinfo {year} {1997})}\BibitemShut
  {NoStop}%
\bibitem [{\citenamefont {Guttman}\ \emph {et~al.}(1998)\citenamefont
  {Guttman}, \citenamefont {Ben-Jacob},\ and\ \citenamefont {Bergman}}]{Gut98}%
  \BibitemOpen
  \bibfield  {author} {\bibinfo {author} {\bibfnamefont {G.~D.}\ \bibnamefont
  {Guttman}}, \bibinfo {author} {\bibfnamefont {E.}~\bibnamefont {Ben-Jacob}},
  \ and\ \bibinfo {author} {\bibfnamefont {D.~J.}\ \bibnamefont {Bergman}},\
  }\href {\doibase 10.1103/PhysRevB.57.2717} {\bibfield  {journal} {\bibinfo
  {journal} {Phys. Rev. B}\ }\textbf {\bibinfo {volume} {57}},\ \bibinfo
  {pages} {2717} (\bibinfo {year} {1998})}\BibitemShut {NoStop}%
\bibitem [{\citenamefont {Scott}\ and\ \citenamefont {Johnson}(1969)}]{Sco69}%
  \BibitemOpen
  \bibfield  {author} {\bibinfo {author} {\bibfnamefont {A.~C.}\ \bibnamefont
  {Scott}}\ and\ \bibinfo {author} {\bibfnamefont {W.~J.}\ \bibnamefont
  {Johnson}},\ }\href {\doibase 10.1063/1.1652665} {\bibfield  {journal}
  {\bibinfo  {journal} {Appl. Phys. Lett.}\ }\textbf {\bibinfo {volume} {14}},\
  \bibinfo {pages} {316} (\bibinfo {year} {1969})}\BibitemShut {NoStop}%
\bibitem [{\citenamefont {Fulton}\ and\ \citenamefont {Dynes}(1973)}]{Ful73}%
  \BibitemOpen
  \bibfield  {author} {\bibinfo {author} {\bibfnamefont {T.}~\bibnamefont
  {Fulton}}\ and\ \bibinfo {author} {\bibfnamefont {R.}~\bibnamefont {Dynes}},\
  }\href {\doibase https://doi.org/10.1016/0038-1098(73)90345-1} {\bibfield
  {journal} {\bibinfo  {journal} {Solid State Commun.}\ }\textbf {\bibinfo
  {volume} {12}},\ \bibinfo {pages} {57 } (\bibinfo {year} {1973})}\BibitemShut
  {NoStop}%
\bibitem [{\citenamefont {Gulevich}\ and\ \citenamefont
  {Kusmartsev}(2007)}]{Gul07}%
  \BibitemOpen
  \bibfield  {author} {\bibinfo {author} {\bibfnamefont {D.~R.}\ \bibnamefont
  {Gulevich}}\ and\ \bibinfo {author} {\bibfnamefont {F.}~\bibnamefont
  {Kusmartsev}},\ }\href@noop {} {\bibfield  {journal} {\bibinfo  {journal}
  {Supercond. Sci. Technol.}\ }\textbf {\bibinfo {volume} {20}},\ \bibinfo
  {pages} {S60} (\bibinfo {year} {2007})}\BibitemShut {NoStop}%
\bibitem [{\citenamefont {Monaco}(2012)}]{Mon12}%
  \BibitemOpen
  \bibfield  {author} {\bibinfo {author} {\bibfnamefont {R.}~\bibnamefont
  {Monaco}},\ }\href@noop {} {\bibfield  {journal} {\bibinfo  {journal}
  {Supercond. Sci. Technol.}\ }\textbf {\bibinfo {volume} {25}},\ \bibinfo
  {pages} {115011} (\bibinfo {year} {2012})}\BibitemShut {NoStop}%
\bibitem [{\citenamefont {Valenti}\ \emph {et~al.}(2014)\citenamefont
  {Valenti}, \citenamefont {Guarcello},\ and\ \citenamefont
  {Spagnolo}}]{Val14}%
  \BibitemOpen
  \bibfield  {author} {\bibinfo {author} {\bibfnamefont {D.}~\bibnamefont
  {Valenti}}, \bibinfo {author} {\bibfnamefont {C.}~\bibnamefont {Guarcello}},
  \ and\ \bibinfo {author} {\bibfnamefont {B.}~\bibnamefont {Spagnolo}},\
  }\href {\doibase 10.1103/PhysRevB.89.214510} {\bibfield  {journal} {\bibinfo
  {journal} {Phys. Rev. B}\ }\textbf {\bibinfo {volume} {89}},\ \bibinfo
  {pages} {214510} (\bibinfo {year} {2014})}\BibitemShut {NoStop}%
\bibitem [{\citenamefont {Zelikman}(2015)}]{Zel15}%
  \BibitemOpen
  \bibfield  {author} {\bibinfo {author} {\bibfnamefont {M.~A.}\ \bibnamefont
  {Zelikman}},\ }\href {\doibase 10.1134/S1063784215090261} {\bibfield
  {journal} {\bibinfo  {journal} {Technical Physics}\ }\textbf {\bibinfo
  {volume} {60}},\ \bibinfo {pages} {1299} (\bibinfo {year}
  {2015})}\BibitemShut {NoStop}%
\bibitem [{\citenamefont {Pankratov}\ \emph {et~al.}(2015)\citenamefont
  {Pankratov}, \citenamefont {Fedorov}, \citenamefont {Salerno}, \citenamefont
  {Shitov},\ and\ \citenamefont {Ustinov}}]{Pan15}%
  \BibitemOpen
  \bibfield  {author} {\bibinfo {author} {\bibfnamefont {A.~L.}\ \bibnamefont
  {Pankratov}}, \bibinfo {author} {\bibfnamefont {K.~G.}\ \bibnamefont
  {Fedorov}}, \bibinfo {author} {\bibfnamefont {M.}~\bibnamefont {Salerno}},
  \bibinfo {author} {\bibfnamefont {S.~V.}\ \bibnamefont {Shitov}}, \ and\
  \bibinfo {author} {\bibfnamefont {A.~V.}\ \bibnamefont {Ustinov}},\ }\href
  {\doibase 10.1103/PhysRevB.92.104501} {\bibfield  {journal} {\bibinfo
  {journal} {Phys. Rev. B}\ }\textbf {\bibinfo {volume} {92}},\ \bibinfo
  {pages} {104501} (\bibinfo {year} {2015})}\BibitemShut {NoStop}%
\bibitem [{\citenamefont {Guarcello}\ \emph
  {et~al.}(2016{\natexlab{b}})\citenamefont {Guarcello}, \citenamefont
  {Valenti}, \citenamefont {Carollo},\ and\ \citenamefont
  {Spagnolo}}]{GuaValSpa16}%
  \BibitemOpen
  \bibfield  {author} {\bibinfo {author} {\bibfnamefont {C.}~\bibnamefont
  {Guarcello}}, \bibinfo {author} {\bibfnamefont {D.}~\bibnamefont {Valenti}},
  \bibinfo {author} {\bibfnamefont {A.}~\bibnamefont {Carollo}}, \ and\
  \bibinfo {author} {\bibfnamefont {B.}~\bibnamefont {Spagnolo}},\ }\href@noop
  {} {\bibfield  {journal} {\bibinfo  {journal} {J. Stat. Mech.: Theory Exp.}\
  }\textbf {\bibinfo {volume} {2016}},\ \bibinfo {pages} {054012} (\bibinfo
  {year} {2016}{\natexlab{b}})}\BibitemShut {NoStop}%
\bibitem [{\citenamefont {Guarcello}\ \emph
  {et~al.}(2017{\natexlab{b}})\citenamefont {Guarcello}, \citenamefont
  {Solinas}, \citenamefont {Di~Ventra},\ and\ \citenamefont
  {Giazotto}}]{GuaSol17}%
  \BibitemOpen
  \bibfield  {author} {\bibinfo {author} {\bibfnamefont {C.}~\bibnamefont
  {Guarcello}}, \bibinfo {author} {\bibfnamefont {P.}~\bibnamefont {Solinas}},
  \bibinfo {author} {\bibfnamefont {M.}~\bibnamefont {Di~Ventra}}, \ and\
  \bibinfo {author} {\bibfnamefont {F.}~\bibnamefont {Giazotto}},\ }\href@noop
  {} {\bibfield  {journal} {\bibinfo  {journal} {Sci. Rep.}\ }\textbf {\bibinfo
  {volume} {7}},\ \bibinfo {pages} {46736} (\bibinfo {year}
  {2017}{\natexlab{b}})}\BibitemShut {NoStop}%
\bibitem [{\citenamefont {Hill}\ \emph {et~al.}(2018)\citenamefont {Hill},
  \citenamefont {Kim},\ and\ \citenamefont {Tserkovnyak}}]{Hil18}%
  \BibitemOpen
  \bibfield  {author} {\bibinfo {author} {\bibfnamefont {D.}~\bibnamefont
  {Hill}}, \bibinfo {author} {\bibfnamefont {S.~K.}\ \bibnamefont {Kim}}, \
  and\ \bibinfo {author} {\bibfnamefont {Y.}~\bibnamefont {Tserkovnyak}},\
  }\href {\doibase 10.1103/PhysRevLett.121.037202} {\bibfield  {journal}
  {\bibinfo  {journal} {Phys. Rev. Lett.}\ }\textbf {\bibinfo {volume} {121}},\
  \bibinfo {pages} {037202} (\bibinfo {year} {2018})}\BibitemShut {NoStop}%
\bibitem [{\citenamefont {Wustmann}\ and\ \citenamefont
  {Osborn}(2018)}]{Wus18}%
  \BibitemOpen
  \bibfield  {author} {\bibinfo {author} {\bibfnamefont {W.}~\bibnamefont
  {Wustmann}}\ and\ \bibinfo {author} {\bibfnamefont {K.~D.}\ \bibnamefont
  {Osborn}},\ }\href@noop {} {\bibfield  {journal} {\bibinfo  {journal} {arXiv
  preprint arXiv:1711.04339}\ } (\bibinfo {year} {2018})}\BibitemShut {NoStop}%
\bibitem [{\citenamefont {Ooi}\ \emph {et~al.}(2007)\citenamefont {Ooi},
  \citenamefont {Savel'ev}, \citenamefont {Gaifullin}, \citenamefont {Mochiku},
  \citenamefont {Hirata},\ and\ \citenamefont {Nori}}]{Ooi07}%
  \BibitemOpen
  \bibfield  {author} {\bibinfo {author} {\bibfnamefont {S.}~\bibnamefont
  {Ooi}}, \bibinfo {author} {\bibfnamefont {S.}~\bibnamefont {Savel'ev}},
  \bibinfo {author} {\bibfnamefont {M.~B.}\ \bibnamefont {Gaifullin}}, \bibinfo
  {author} {\bibfnamefont {T.}~\bibnamefont {Mochiku}}, \bibinfo {author}
  {\bibfnamefont {K.}~\bibnamefont {Hirata}}, \ and\ \bibinfo {author}
  {\bibfnamefont {F.}~\bibnamefont {Nori}},\ }\href {\doibase
  10.1103/PhysRevLett.99.207003} {\bibfield  {journal} {\bibinfo  {journal}
  {Phys. Rev. Lett.}\ }\textbf {\bibinfo {volume} {99}},\ \bibinfo {pages}
  {207003} (\bibinfo {year} {2007})}\BibitemShut {NoStop}%
\bibitem [{\citenamefont {Likharev}(2012)}]{Lik12}%
  \BibitemOpen
  \bibfield  {author} {\bibinfo {author} {\bibfnamefont {K.~K.}\ \bibnamefont
  {Likharev}},\ }\href {\doibase https://doi.org/10.1016/j.physc.2012.05.016}
  {\bibfield  {journal} {\bibinfo  {journal} {Physica (Amsterdam)}\ }\textbf
  {\bibinfo {volume} {482C}},\ \bibinfo {pages} {6 } (\bibinfo {year}
  {2012})}\BibitemShut {NoStop}%
\bibitem [{\citenamefont {Fedorov}\ \emph {et~al.}(2012)\citenamefont
  {Fedorov}, \citenamefont {Shitov}, \citenamefont {Rotzinger},\ and\
  \citenamefont {Ustinov}}]{Fed12}%
  \BibitemOpen
  \bibfield  {author} {\bibinfo {author} {\bibfnamefont {K.~G.}\ \bibnamefont
  {Fedorov}}, \bibinfo {author} {\bibfnamefont {S.~V.}\ \bibnamefont {Shitov}},
  \bibinfo {author} {\bibfnamefont {H.}~\bibnamefont {Rotzinger}}, \ and\
  \bibinfo {author} {\bibfnamefont {A.~V.}\ \bibnamefont {Ustinov}},\ }\href
  {\doibase 10.1103/PhysRevB.85.184512} {\bibfield  {journal} {\bibinfo
  {journal} {Phys. Rev. B}\ }\textbf {\bibinfo {volume} {85}},\ \bibinfo
  {pages} {184512} (\bibinfo {year} {2012})}\BibitemShut {NoStop}%
\bibitem [{\citenamefont {Monaco}\ \emph {et~al.}(2013)\citenamefont {Monaco},
  \citenamefont {Granata}, \citenamefont {Russo},\ and\ \citenamefont
  {Vettoliere}}]{Mon13}%
  \BibitemOpen
  \bibfield  {author} {\bibinfo {author} {\bibfnamefont {R.}~\bibnamefont
  {Monaco}}, \bibinfo {author} {\bibfnamefont {C.}~\bibnamefont {Granata}},
  \bibinfo {author} {\bibfnamefont {R.}~\bibnamefont {Russo}}, \ and\ \bibinfo
  {author} {\bibfnamefont {A.}~\bibnamefont {Vettoliere}},\ }\href@noop {}
  {\bibfield  {journal} {\bibinfo  {journal} {Supercond. Sci. Technol.}\
  }\textbf {\bibinfo {volume} {26}},\ \bibinfo {pages} {125005} (\bibinfo
  {year} {2013})}\BibitemShut {NoStop}%
\bibitem [{\citenamefont {Granata}\ \emph {et~al.}(2014)\citenamefont
  {Granata}, \citenamefont {Vettoliere},\ and\ \citenamefont {Monaco}}]{Gra14}%
  \BibitemOpen
  \bibfield  {author} {\bibinfo {author} {\bibfnamefont {C.}~\bibnamefont
  {Granata}}, \bibinfo {author} {\bibfnamefont {A.}~\bibnamefont {Vettoliere}},
  \ and\ \bibinfo {author} {\bibfnamefont {R.}~\bibnamefont {Monaco}},\
  }\href@noop {} {\bibfield  {journal} {\bibinfo  {journal} {Supercond. Sci.
  Technol.}\ }\textbf {\bibinfo {volume} {27}},\ \bibinfo {pages} {095003}
  (\bibinfo {year} {2014})}\BibitemShut {NoStop}%
\bibitem [{\citenamefont {Koshelets}(2014)}]{Kos14}%
  \BibitemOpen
  \bibfield  {author} {\bibinfo {author} {\bibfnamefont {V.~P.}\ \bibnamefont
  {Koshelets}},\ }\href@noop {} {\bibfield  {journal} {\bibinfo  {journal}
  {Supercond. Sci. Technol.}\ }\textbf {\bibinfo {volume} {27}},\ \bibinfo
  {pages} {065010} (\bibinfo {year} {2014})}\BibitemShut {NoStop}%
\bibitem [{\citenamefont {Fedorov}\ \emph {et~al.}(2014)\citenamefont
  {Fedorov}, \citenamefont {Shcherbakova}, \citenamefont {Wolf}, \citenamefont
  {Beckmann},\ and\ \citenamefont {Ustinov}}]{Fed14}%
  \BibitemOpen
  \bibfield  {author} {\bibinfo {author} {\bibfnamefont {K.~G.}\ \bibnamefont
  {Fedorov}}, \bibinfo {author} {\bibfnamefont {A.~V.}\ \bibnamefont
  {Shcherbakova}}, \bibinfo {author} {\bibfnamefont {M.~J.}\ \bibnamefont
  {Wolf}}, \bibinfo {author} {\bibfnamefont {D.}~\bibnamefont {Beckmann}}, \
  and\ \bibinfo {author} {\bibfnamefont {A.~V.}\ \bibnamefont {Ustinov}},\
  }\href {\doibase 10.1103/PhysRevLett.112.160502} {\bibfield  {journal}
  {\bibinfo  {journal} {Phys. Rev. Lett.}\ }\textbf {\bibinfo {volume} {112}},\
  \bibinfo {pages} {160502} (\bibinfo {year} {2014})}\BibitemShut {NoStop}%
\bibitem [{\citenamefont {Vettoliere}\ \emph {et~al.}(2015)\citenamefont
  {Vettoliere}, \citenamefont {Granata},\ and\ \citenamefont {Monaco}}]{Vet15}%
  \BibitemOpen
  \bibfield  {author} {\bibinfo {author} {\bibfnamefont {A.}~\bibnamefont
  {Vettoliere}}, \bibinfo {author} {\bibfnamefont {C.}~\bibnamefont {Granata}},
  \ and\ \bibinfo {author} {\bibfnamefont {R.}~\bibnamefont {Monaco}},\ }\href
  {\doibase 10.1109/TMAG.2014.2357473} {\bibfield  {journal} {\bibinfo
  {journal} {IEEE Trans. Magn.}\ }\textbf {\bibinfo {volume} {51}},\ \bibinfo
  {pages} {1} (\bibinfo {year} {2015})}\BibitemShut {NoStop}%
\bibitem [{\citenamefont {Golovchanskiy}\ \emph {et~al.}(2017)\citenamefont
  {Golovchanskiy}, \citenamefont {Abramov}, \citenamefont {Stolyarov},
  \citenamefont {Emelyanova}, \citenamefont {Golubov}, \citenamefont
  {Ustinov},\ and\ \citenamefont {Ryazanov}}]{Gol17}%
  \BibitemOpen
  \bibfield  {author} {\bibinfo {author} {\bibfnamefont {I.~A.}\ \bibnamefont
  {Golovchanskiy}}, \bibinfo {author} {\bibfnamefont {N.~N.}\ \bibnamefont
  {Abramov}}, \bibinfo {author} {\bibfnamefont {V.~S.}\ \bibnamefont
  {Stolyarov}}, \bibinfo {author} {\bibfnamefont {O.~V.}\ \bibnamefont
  {Emelyanova}}, \bibinfo {author} {\bibfnamefont {A.~A.}\ \bibnamefont
  {Golubov}}, \bibinfo {author} {\bibfnamefont {A.~V.}\ \bibnamefont
  {Ustinov}}, \ and\ \bibinfo {author} {\bibfnamefont {V.~V.}\ \bibnamefont
  {Ryazanov}},\ }\href@noop {} {\bibfield  {journal} {\bibinfo  {journal}
  {Supercond. Sci. Technol.}\ }\textbf {\bibinfo {volume} {30}},\ \bibinfo
  {pages} {054005} (\bibinfo {year} {2017})}\BibitemShut {NoStop}%
\bibitem [{\citenamefont {Davidson}\ \emph {et~al.}(1986)\citenamefont
  {Davidson}, \citenamefont {Dueholm},\ and\ \citenamefont {Pedersen}}]{Dav86}%
  \BibitemOpen
  \bibfield  {author} {\bibinfo {author} {\bibfnamefont {A.}~\bibnamefont
  {Davidson}}, \bibinfo {author} {\bibfnamefont {B.}~\bibnamefont {Dueholm}}, \
  and\ \bibinfo {author} {\bibfnamefont {N.~F.}\ \bibnamefont {Pedersen}},\
  }\href {\doibase 10.1063/1.337324} {\bibfield  {journal} {\bibinfo  {journal}
  {, J. Appl. Phys.}\ }\textbf {\bibinfo {volume} {60}},\ \bibinfo {pages}
  {1447} (\bibinfo {year} {1986})}\BibitemShut {NoStop}%
\bibitem [{\citenamefont {Ustinov}\ \emph {et~al.}(1992)\citenamefont
  {Ustinov}, \citenamefont {Doderer}, \citenamefont {Huebener}, \citenamefont
  {Pedersen}, \citenamefont {Mayer},\ and\ \citenamefont {Oboznov}}]{Ust92}%
  \BibitemOpen
  \bibfield  {author} {\bibinfo {author} {\bibfnamefont {A.~V.}\ \bibnamefont
  {Ustinov}}, \bibinfo {author} {\bibfnamefont {T.}~\bibnamefont {Doderer}},
  \bibinfo {author} {\bibfnamefont {R.~P.}\ \bibnamefont {Huebener}}, \bibinfo
  {author} {\bibfnamefont {N.~F.}\ \bibnamefont {Pedersen}}, \bibinfo {author}
  {\bibfnamefont {B.}~\bibnamefont {Mayer}}, \ and\ \bibinfo {author}
  {\bibfnamefont {V.~A.}\ \bibnamefont {Oboznov}},\ }\href {\doibase
  10.1103/PhysRevLett.69.1815} {\bibfield  {journal} {\bibinfo  {journal}
  {Phys. Rev. Lett.}\ }\textbf {\bibinfo {volume} {69}},\ \bibinfo {pages}
  {1815} (\bibinfo {year} {1992})}\BibitemShut {NoStop}%
\bibitem [{\citenamefont {Ustinov}(2002)}]{Ust02}%
  \BibitemOpen
  \bibfield  {author} {\bibinfo {author} {\bibfnamefont {A.~V.}\ \bibnamefont
  {Ustinov}},\ }\href {\doibase 10.1063/1.1474617} {\bibfield  {journal}
  {\bibinfo  {journal} {Appl. Phys. Lett.}\ }\textbf {\bibinfo {volume} {80}},\
  \bibinfo {pages} {3153} (\bibinfo {year} {2002})}\BibitemShut {NoStop}%
\bibitem [{\citenamefont {Harris}(1974)}]{Har74}%
  \BibitemOpen
  \bibfield  {author} {\bibinfo {author} {\bibfnamefont {R.~E.}\ \bibnamefont
  {Harris}},\ }\href {\doibase 10.1103/PhysRevB.10.84} {\bibfield  {journal}
  {\bibinfo  {journal} {Phys. Rev. B}\ }\textbf {\bibinfo {volume} {10}},\
  \bibinfo {pages} {84} (\bibinfo {year} {1974})}\BibitemShut {NoStop}%
\bibitem [{\citenamefont {Guarcello}\ \emph
  {et~al.}(2018{\natexlab{d}})\citenamefont {Guarcello}, \citenamefont
  {Braggio}, \citenamefont {Solinas},\ and\ \citenamefont
  {Giazotto}}]{Gua18bis}%
  \BibitemOpen
  \bibfield  {author} {\bibinfo {author} {\bibfnamefont {C.}~\bibnamefont
  {Guarcello}}, \bibinfo {author} {\bibfnamefont {A.}~\bibnamefont {Braggio}},
  \bibinfo {author} {\bibfnamefont {P.}~\bibnamefont {Solinas}}, \ and\
  \bibinfo {author} {\bibfnamefont {F.}~\bibnamefont {Giazotto}},\ }\href@noop
  {} {\bibfield  {journal} {\bibinfo  {journal} {arXiv preprint
  arXiv:1807.03186}\ } (\bibinfo {year} {2018}{\natexlab{d}})}\BibitemShut
  {NoStop}%
\bibitem [{\citenamefont {Barends}\ \emph {et~al.}(2008)\citenamefont
  {Barends}, \citenamefont {Baselmans}, \citenamefont {Yates}, \citenamefont
  {Gao}, \citenamefont {Hovenier},\ and\ \citenamefont {Klapwijk}}]{Bar08}%
  \BibitemOpen
  \bibfield  {author} {\bibinfo {author} {\bibfnamefont {R.}~\bibnamefont
  {Barends}}, \bibinfo {author} {\bibfnamefont {J.~J.~A.}\ \bibnamefont
  {Baselmans}}, \bibinfo {author} {\bibfnamefont {S.~J.~C.}\ \bibnamefont
  {Yates}}, \bibinfo {author} {\bibfnamefont {J.~R.}\ \bibnamefont {Gao}},
  \bibinfo {author} {\bibfnamefont {J.~N.}\ \bibnamefont {Hovenier}}, \ and\
  \bibinfo {author} {\bibfnamefont {T.~M.}\ \bibnamefont {Klapwijk}},\ }\href
  {\doibase 10.1103/PhysRevLett.100.257002} {\bibfield  {journal} {\bibinfo
  {journal} {Phys. Rev. Lett.}\ }\textbf {\bibinfo {volume} {100}},\ \bibinfo
  {pages} {257002} (\bibinfo {year} {2008})}\BibitemShut {NoStop}%
\bibitem [{\citenamefont {Kaplan}\ \emph {et~al.}(1976)\citenamefont {Kaplan},
  \citenamefont {Chi}, \citenamefont {Langenberg}, \citenamefont {Chang},
  \citenamefont {Jafarey},\ and\ \citenamefont {Scalapino}}]{Kap76}%
  \BibitemOpen
  \bibfield  {author} {\bibinfo {author} {\bibfnamefont {S.~B.}\ \bibnamefont
  {Kaplan}}, \bibinfo {author} {\bibfnamefont {C.~C.}\ \bibnamefont {Chi}},
  \bibinfo {author} {\bibfnamefont {D.~N.}\ \bibnamefont {Langenberg}},
  \bibinfo {author} {\bibfnamefont {J.~J.}\ \bibnamefont {Chang}}, \bibinfo
  {author} {\bibfnamefont {S.}~\bibnamefont {Jafarey}}, \ and\ \bibinfo
  {author} {\bibfnamefont {D.~J.}\ \bibnamefont {Scalapino}},\ }\href {\doibase
  10.1103/PhysRevB.14.4854} {\bibfield  {journal} {\bibinfo  {journal} {Phys.
  Rev. B}\ }\textbf {\bibinfo {volume} {14}},\ \bibinfo {pages} {4854}
  (\bibinfo {year} {1976})}\BibitemShut {NoStop}%
\bibitem [{\citenamefont {Luomahaara}\ \emph {et~al.}(2014)\citenamefont
  {Luomahaara}, \citenamefont {Vesterinen}, \citenamefont {Gr{\"o}nberg},\ and\
  \citenamefont {Hassel}}]{Luo14}%
  \BibitemOpen
  \bibfield  {author} {\bibinfo {author} {\bibfnamefont {J.}~\bibnamefont
  {Luomahaara}}, \bibinfo {author} {\bibfnamefont {V.}~\bibnamefont
  {Vesterinen}}, \bibinfo {author} {\bibfnamefont {L.}~\bibnamefont
  {Gr{\"o}nberg}}, \ and\ \bibinfo {author} {\bibfnamefont {J.}~\bibnamefont
  {Hassel}},\ }\href@noop {} {\bibfield  {journal} {\bibinfo  {journal} {Nat.
  Commun.}\ }\textbf {\bibinfo {volume} {5}},\ \bibinfo {pages} {4872}
  (\bibinfo {year} {2014})}\BibitemShut {NoStop}%
\bibitem [{\citenamefont {Dynes}\ \emph {et~al.}(1978)\citenamefont {Dynes},
  \citenamefont {Narayanamurti},\ and\ \citenamefont {Garno}}]{Dyn78}%
  \BibitemOpen
  \bibfield  {author} {\bibinfo {author} {\bibfnamefont {R.~C.}\ \bibnamefont
  {Dynes}}, \bibinfo {author} {\bibfnamefont {V.}~\bibnamefont
  {Narayanamurti}}, \ and\ \bibinfo {author} {\bibfnamefont {J.~P.}\
  \bibnamefont {Garno}},\ }\href {\doibase 10.1103/PhysRevLett.41.1509}
  {\bibfield  {journal} {\bibinfo  {journal} {Phys. Rev. Lett.}\ }\textbf
  {\bibinfo {volume} {41}},\ \bibinfo {pages} {1509} (\bibinfo {year}
  {1978})}\BibitemShut {NoStop}%
\end{thebibliography}

%

\end{document}